\documentclass[12pt]{article}
\usepackage{amssymb}
\usepackage{latexsym}
\usepackage{amsmath}
\usepackage{color}
\usepackage{cite}
\usepackage{units}
\usepackage{tabularx}
\usepackage{graphicx}
\usepackage{cancel}

\numberwithin{equation}{section}

\begin{document}
\renewcommand{\thefootnote}{\fnsymbol{footnote} }

\pagestyle{empty}
\begin{flushright}
July 2013
\end{flushright}

\begin{center}
{\large\sc {\bf Determination of the CMSSM Parameters using Neural Networks}}  

\vspace{1cm}
{\sc Nicki Bornhauser$^{1}$ and Manuel Drees$^{1}$}  

\vspace*{5mm}
{}$^1${\it Physikalisches Institut and Bethe Center for Theoretical
  Physics, Universit\"at Bonn,\\ Nu\ss{}allee 12, D--53115 Bonn, Germany} 
\end{center}

\vspace*{1cm}
\begin{abstract}
  In most (weakly interacting) extensions of the Standard Model the
  relation mapping the parameter values onto experimentally measurable
  quantities can be computed (with some uncertainties), but the
  inverse relation is usually not known. In this paper we demonstrate
  the ability of artificial neural networks to find this unknown
  relation, by determining the unknown parameters of the constrained
  minimal supersymmetric extension of the Standard Model (CMSSM) from
  quantities that can be measured at the LHC. We expect that the
  method works also for many other new physics models. We compare its
  performance with the results of a straightforward $\chi^2$
  minimization. We simulate LHC signals at a center of mass energy of
  $\unit[14]{TeV}$ at the hadron level. In this proof--of--concept
  study we do not explicitly simulate Standard Model backgrounds, but
  apply cuts that have been shown to enhance the
  signal--to--background ratio. We analyze four different benchmark
  points that lie just beyond current lower limits on superparticle
  masses, each of which leads to around $1000$ events after cuts for
  an integrated luminosity of $\unit[10]{fb^{-1}}$. We use up to $84$
  observables, most of which are counting observables; we do not
  attempt to directly reconstruct (differences of) masses from
  kinematic edges or kinks of distributions. We nevertheless find that
  $m_0$ and $m_{1/2}$ can be determined reliably, with errors as small
  as $1\%$ in some cases. With $\unit[500]{fb^{-1}}$ of data
  $\tan\beta$ as well as $A_0$ can also be determined quite
  accurately. For comparable computational effort the $\chi^2$
  minimization yielded much worse results.
\end{abstract}

\newpage
\setcounter{page}{1}
\pagestyle{plain}

\section{Introduction}
\label{sec:Introduction}

The Large Hadron Collider (LHC) is running successfully. After the
next long shutdown the center of mass energy will be raised from 8 to
13 or 14 TeV. This higher center of mass energy will increase the
reach for finding new physics. Here we are concerned with
supersymmetric extensions of the Standard Model (SM) of particle
physics. Within the simplest potentially realistic supersymmetric
model, the minimal supersymmetric extension of the SM (MSSM)
\cite{mssm}, the 14 TeV LHC will increase the mass reach for first
generation squarks and gluinos from the current lower bounds, which
reach nearly 1.5 TeV for equal squark and gluino masses
\cite{bounds},\footnote{Mass limits are usually cited for the
  constrained MSSM (CMSSM) or for certain simplified scenarios; even
  in these scenarios the gluino mass bound reduces to about 900 GeV if
  squarks are much heavier than gluinos. In the more general MSSM the
  bounds may be somewhat weaker \cite{kraml}; however, they can
  be reduced significantly only if strongly interacting sparticles are
  quite close in mass to the lightest superparticle \cite{jamie},
  which is not particularly natural from the model building point of
  view.} to values between 2 and 3 TeV \cite{reach}. This leaves
plenty of room for new discoveries. In particular, the ``natural''
range of parameters of supersymmetric theories will then be probed
decisively. 

Discovering a signal for physics beyond the SM, important as it would
be, would certainly not be the end of the LHC physics program. One
would then not only have to ascertain what kind of new physics has
been discovered, but also determine the values of the free parameters
as accurately as possible. In the context of the MSSM, this should
help to unravel the mechanism responsible for the breaking of
supersymmetry.

There is a large literature on ways to determine the parameters of
supersymmetric theories. Most methods start from kinematic features,
in particular endpoints or ``edges'' of invariant mass distributions
\cite{edges} or kinks in slightly more complicated kinematic
distributions \cite{kinks}. These kinematic features directly allow to
determine (differences of) superparticle masses; at least at the
tree--level, in most cases there is a direct relation between the mass
of a superpartner and a weak--scale parameter of the underlying
theory. In many cases the experimental resolution that can be achieved
is such that at least one--loop corrections should be included;
e.g.~the difference between pole (on--shell) masses, which determine
the kinematics, and $\overline{\rm DR}$ masses, which are
``fundamental'' free parameters in the supersymmetric Les Houches
accord \cite{slha}, can easily reach several percent for strongly
interacting superparticles \cite{loopmass}. The derived $\overline{\rm
  DR}$ masses will then depend on many (pole) masses. Moreover, in the
chargino and neutralino sector, as well as for third generation
sfermions, the relation between pole masses and fundamental
parameters is complicated by mixing \cite{mssm}. Nevertheless the
basic kinematic quantities that are used for parameter determination
can be determined from a single (simulated) experiment. While the step
from there to the determination of the basic parameters and their
errors may entail many calls of spectrum calculators, there is no need
to simulate event generation for different sets of parameters, which
is usually far more time consuming than the calculation of the
spectrum of superparticles.\footnote{This statement may no longer be
  strictly true in the presence of additional hard radiation, which
  changes the shapes of kinematic distributions near the features
  (edges or kinks) used to determine masses \cite{radiation}. Beyond
  the collinear approximation, the amount of radiation emitted may
  depend on (combinations of) superparticle masses in a complicated
  manner. However, in practice the shapes of the relevant kinematic
  distributions are fitted to (simulated) data rather than directly
  taken from theory. While we are not aware of an analysis of
  kinematical fitting based on a fully NLO corrected event simulation,
  we expect that by fitting the shapes of the relevant distributions
  one can still determine the locations of the relevant kinematic
  features without prior knowledge of superparticle masses.}

On the other hand, even in constrained scenarios purely kinematical
determinations of the underlying parameters work well only if
sufficiently many events contain two (or more) charged leptons
(meaning electrons or muons). Kinematic reconstructions based on jets
suffer not only from the much poorer energy resolution of jets, but
also from larger combinatorial backgrounds (since the production and
decay of strongly interacting superparticles typically leads to events
with many jets).

In any case, it is clear that the number of signal events in certain
categories contains a lot of information about the underlying
physics. Even if kinematic reconstruction works well, it would be
wasteful to ignore this information. To mention a well--known example,
the cross section for the pair production of a new color triplet
complex scalar boson (like the stop) is much smaller than that for
spin$-1/2$ quarks of the same mass \cite{mssm}. Moreover, in
constrained supersymmetric scenarios strongly interacting
superparticles tend to be heavier than those without strong
interactions. The production of strongly interacting superparticles
therefore frequently leads to long ``cascade'' decays \cite{cascade},
which can populate many ``topologically different'' final states,
i.e.~final states characterized by different numbers (and charges) of
leptons as well as different numbers (and flavors) of jets. It has
been recognized quite early that the relative abundance of these final
states contains a great deal of information about the sparticle
spectrum \cite{early_rate}.

However, these early studies mostly focused on distinguishing
qualitatively different spectra of superparticles. Information on the
total signal rate has only quite recently been included in fits
attempting to determine the underlying parameters from (simulated)
events \cite{combined}. We are not aware of any study that attempts to
determine the values of the underlying parameters using (mostly)
counting observables, although a recent analysis showed that these
observables can be very useful for discriminating between discrete
sets of model parameters \cite{bod4}. One major difficulty with this
approach is that it requires to generate event samples for many
different assumed sets of input parameters. For example, even in the
CMSSM, which has only four free parameters, a simple grid scan over
all parameters with a step size comparable to or smaller than the
anticipated statistical accuracy of the method is prohibitively CPU
expensive in most circumstances.

The purpose of this paper is to demonstrate the usefulness of
artificial neural networks for the parameter determination of such a
new physics theory. Largely due to the limitation of our computational
resources, we do this in the framework of the CMSSM; the method should
also be useful for other theories, supersymmetric or otherwise. Again
for computational simplicity we ignore detector effects, but we work
at full hadron level, including initial and final state radiation,
hadronization, and the underlying event. Similarly, we ignore Standard
Model backgrounds, but we include cuts that should keep them at a
manageable level. We consider four benchmark scenarios, all of which
lie (slightly) beyond current lower bounds on superparticle masses,
but have qualitatively different spectra. We find that $10 \ {\rm
  fb}^{-1}$ of data at $\sqrt{s} = 14 \ {\rm TeV}$ are sufficient to
determine the common scalar mass parameter $m_0$ and the common
gaugino mass parameter $m_{1/2}$ to few percent accuracy {\em without}
any direct kinematic mass reconstruction. With $500 \ {\rm fb}^{-1}$
of data the neural networks can also determine the trilinear soft
breaking parameter $A_0$ and the ratio of vacuum expectation values
$\tan\beta$ quite accurately for these benchmark scenarios. In
contrast, in many cases a simple $\chi^2$ minimization failed to
converge, i.e.~it could not reliably determine the parameters and
their errors. The likely reason is that the minimization of $\chi^2$
is very sensitive to fluctuations in the predictions due to finite
Monte Carlo statistics.

In this paper we are only interested in the production and decay of
superparticles at the LHC, as an example for an extension of the SM
containing many new parameters that can hopefully be determined from
future LHC data. In our numerical analysis we will therefore respect
the experimental bounds on the masses of superparticles, but we will
not try to reproduce the recently discovered (increasingly)
Higgs--like boson \cite{higgs_dis} in our CMSSM spectra, nor will we
try to describe Dark Matter through thermally produced
superparticles. Instead we are using the CMSSM as toy model whose
parameter space is manageable even without requiring the correct Higgs
mass and Dark Matter relic density. Obviously imposing these
constraints, or other constraints not directly related to LHC data,
would simplify the task of fixing the free parameters. Here we wish to
show that data on the production and decay of superparticles at the
LHC by themselves can be used for this task, even if mostly counting
observables are used.

The remainder of this article is organized as follows.  In
Sec.~\ref{sec:Simulation} we first introduce the general setting of
the simulation. One important issue is the choice of observables. An
automated reconstruction of the underlying parameters can only succeed
if one has sufficiently many observables to be sensitive to all
parameters everywhere in parameter space. On the other hand, including
too many observables can dilute the statistical power. We present a
set of observables which we showed to be useful for discriminating
between different parameter sets for a more general supersymmetric
model with 15 parameters \cite{bod4}. In the second part of
Sec.~\ref{sec:Simulation} we introduce four different benchmark points
in the CMSSM framework. In Sec.~\ref{sec:Strategies} we discuss both
our attempts at parameter reconstruction, first using artificial
neural networks and second a $\chi^2$ minimization. We explain the
general set--up as well as each step of the creation of the neural
networks for this specific application. We also estimate the errors on
the CMSSM parameters, including their correlations, using different
methods that yield consistent results. In the second part of
Sec.~\ref{sec:Strategies} the attempted $\chi^2$ minimization is
discussed; as already mentioned, it does not perform very well. The
results obtained by the artificial neural networks for all four
benchmark points are discussed in Sec.~\ref{sec:Results}. We also
compare them to the results from the $\chi^2$ minimization. Finally,
the last Section contains a summary and some conclusions.

\section{Simulation}
\label{sec:Simulation}

We simulate future LHC data at a center of mass energy of
$\unit[14]{TeV}$. As mentioned in the Introduction, we work in the
framework of the CMSSM, where the entire spectrum of superparticles
and Higgs bosons is defined by four continuous parameters and a
sign. The continuous parameters are the common scalar mass parameter
$m_0$, the common gaugino mass $m_{1/2}$, the common trilinear soft
breaking parameter $A_0$ and the ratio $\tan\beta$ of vacuum
expectation values of the two Higgs doublets. As usual, $m_0, \,
m_{1/2}$ and $A_0$ are specified at the scale of Grand Unification,
$M_X \simeq 2 \cdot 10^{16}$ GeV, whereas $\tan\beta$ is given at the
electroweak scale. We fix the sign of the supersymmetric higgsino mass
parameter $\mu$ to be positive.

We use SOFTSUSY \cite{bib:softsusy} to compute the CMSSM superparticle
and Higgs boson spectra from the values of the four input
parameters. The weak--scale spectrum is then passed on to SUSY--HIT
\cite{bib:susyhit}, which calculates the branching ratios of all
kinematically allowed decays. Knowledge of the superparticle masses
and branching ratios is needed for the simulation of the production
and decay of pairs of superparticles at the LHC, which is handled by
the event generator Herwig++ \cite{bib:herwig}. In a first step
$10,000$ events are simulated in order to determine the total cross
section for the production of all superparticles for the given set of
input parameters. Next, the appropriate number of events is simulated
which corresponds to the assumed integrated luminosity; we will show
results for $10$ and $500 \ {\rm fb}^{-1}$. 

Each simulated event is assigned to one of twelve mutually exclusive
event classes, based on the number, charges and flavors of charged
leptons. In addition, for each event a small number of mostly counting
observables is kept, from which we construct our $84$ observables. This is
described in more detail in the following Subsection. We do this,
first of all, for four benchmark scenarios, which lie in qualitatively
different regions of CMSSM parameter space, as described in the second
Subsection. Of course, in the attempt to determine the values of the
CMSSM parameters from the four simulated measurements, the procedure
from spectrum calculation to event generation has to be performed for
many additional parameter sets, as described in Sec.~\ref{sec:Strategies}.

\subsection{Observables and their Covariances}
\label{sec:Observables}

In this Subsection we summarize our observables, which we introduced
in detail in Sec.~3 of \cite{bod4}. In particular, the precise
definitions of the objects (isolated charged leptons, hadronically
decaying $\tau$ leptons, hadronic jets with or without $b-$tag) we
use to characterize the events, and the applied cuts, can be found in
the Appendices of \cite{bod4}. 

As already noted, we group all accepted events into twelve mutually
exclusive classes, which differ by the number, charges and flavors of
charged leptons. Here only isolated electrons or muons with transverse
momentum $p_T > 10$ GeV and pseudorapidity $|\eta| < 2.5$ are
counted. These classes are:
\begin{itemize}
\item[1.] $0l$: Events with no charged leptons
\item[2.] $1l^-$: Events with exactly one charged lepton, with
  negative charge (in units of the proton charge)
\item[3.] $1l^+$: Events with exactly one charged lepton, with
  positive charge
\item[4.] $2l^-$: Events with exactly two charged leptons, with total
  charge $-2$
\item[5.] $2l^+$: Events with exactly two charged leptons, with total
  charge $+2$
\item[6.] $l^+_i l^-_i$: Events with exactly two charged leptons, with
  opposite charge but the same flavor; i.e.~$e^- e^+$ or $\mu^+ \mu^-$
\item[7.] $l^+_i l^-_{j;\ j \neq i}$: Events with exactly two
  charged leptons, with opposite charge and different flavor; i.e.~$e^-
  \mu^+$ or $e^+ \mu^-$
\item[8.] $l^-_i l^-_j l^+_j$: Events with exactly three charged
  leptons with total charge $-1$. There is an opposite--charged
  lepton pair with same flavor. For example $e^- \mu^- \mu^+$ or $e^-
  e^- e^+$ 
\item[9.] $l^+_i l^+_j l^-_j$: Events with exactly three charged leptons with
  total charge $+1$. There is an opposite--charged lepton pair with
  same flavor. For example $e^+ \mu^- \mu^+$ or $e^+e^-e^+$
\item[10.] $l^-_i l^-_j l^{\pm}_{k;\ k \neq j, i \ {\rm for} \ +}$: Events
  with exactly three charged leptons with total negative charge,
  i.e.~there are at least two negatively charged leptons. There is
  \underline{no} opposite--charged lepton pair with same flavor. For
  example $e^- e^- \mu^+$ or $e^-e^-e^-$ 
\item[11.] $l^+_i l^+_j l^{\pm}_{k;\ k \neq j, i \ {\rm for} \ -}$: Events
  with exactly three charged leptons with total positive charge,
  i.e.~there are at least two positively charged leptons. There is
  \underline{no} opposite--charged lepton pair with same flavor. For
  example $e^+ e^+ \mu^-$ or $e^+e^+e^+$
\item[12.] $4l$: Events with four or more charged leptons
\end{itemize}
We distinguish between different charges of leptons since the initial
state at the LHC carries charge $+2$. In general the number of events
with positively charged leptons can therefore differ from those with
negatively charged leptons. Moreover, we distinguish between lepton
pairs with opposite charge but the same flavor, which can originate
from leptonic neutralino decays, $\tilde \chi_a^0 \rightarrow
l_i^+l_i^- \tilde \chi_b^0$, and all other lepton pairs, which have to
come from the decays of two different particles. This explains why we
have two different classes of events with exactly one charged lepton,
and four different classes each for events with exactly two and
exactly three charged leptons, respectively. In principle we could
also define several different classes of four lepton events. However,
the number of such events is in any case rather small; further
separating these few events into several classes is therefore not very
useful.\footnote{In fact, separating the events into too many distinct
  classes is harmful. The reason is that we wish to use Gaussian
  statistics; our observables become approximately Gaussian only
  in the limit of large event numbers. We therefore only use event
  classes containing some minimal number of events, as described in
  Sec.~3.}

Our first observable is the total number of events after cuts,
$N$. Note that the cuts differ for the different event classes, as
described in ref.\cite{bod4}. In addition, for each of these twelve
classes $c \in \{1, 2, \dots, 12\}$, the values of seven observables
$O_{i,c}, \ i \in \{1, 2, \dots, 7\}$ are computed:
\begin{itemize}
\item $O_{1,c} = n_c/N$: The number of events $n_c$ contained in the
  given class $c$ divided by the total number of events $N$, i.e.~the
  fraction of all events contained in a given class
\item $O_{2,c} = \langle \tau^- \rangle_c$: Average number of tagged
  hadronically decaying $\tau^-$ of all events within a given class
  $c$ 
\item $O_{3,c} = \langle \tau^+ \rangle_c$: Average number of tagged
  hadronically decaying $\tau^+$ of all events within a given class
  $c$ 
\item $O_{4,c} = \langle b \rangle_c$: Average number of tagged
  $b-$jets of all events within a given class $c$ 
\item $O_{5,c} = \langle j \rangle_c$: Average number of non$-b-$jets
  of all events within a given class $c$
\item $O_{6,c} = \langle j^2 \rangle_c$: Average of the square of the
  number of non$-b-$jets\footnote{If event $i$ in the given class
    contains $N_j^{(i)}$ non--$b-$jets, then $\langle j^2 \rangle_c =
    1/n_c \sum_{i=1}^{n_c} \left( N_j^{(i)} \right)^2$.} of all events
  within a given class $c$
\item $O_{7,c} = \langle H_T \rangle_c$: Average value of $H_T$ of all
events within a given class $c$, where $H_T$ is the scalar sum of the
transverse momenta of all hard objects, including the missing $p_T$
\end{itemize}
Both $\tau-$ and $b-$jets have to have transverse momentum $p_T > 20$
GeV and pseudorapidity $|\eta| < 2.5$. In addition, a $\tau-$jet needs
to be isolated, and a $b-$jet has to contain a $b-$flavored
hadron. Jets satisfying these criteria are tagged with an assumed
tagging efficiency of $\unit[50]{\%}$. Finally, $H_T$ is the scalar
sum of the transverse momenta of all hard objects (jets and charged
leptons) and the absolute value of the missing $p_T$. We again refer
to ref.\cite{bod4} for further details.

Three of those observables are different to the ones used in
ref.\cite{bod4}. The number of events in a given class $c$ that
contain at least one tagged hadronically decaying $\tau^-$ divided by
the total number of events in this class, $n_{c,\tau^-}/n_c$, has been
replaced by the average number of tagged hadronically decaying
$\tau^-$ of all events within a given class $c$, $\langle \tau^-
\rangle_c$, and similarly for positively charged $\tau-$jets. In the
parameter sets considered in \cite{bod4} the number of events
containing a tagged $\tau$ was rather small and the number of events
containing two of those even smaller. Therefore it was sufficient to
just count the number of events containing at least one tagged
$\tau$. Now in the case of the CMSSM there can be more events with a
higher number of $\tau-$leptons. Therefore here we switched the
observable to preserve more information about the measurement. The
same applies to the observable $\langle b \rangle_c$, which is used
instead of $n_{c,b}/n_c$.

Out of the $85$ observables listed above, one should be discarded.
Obviously the fractions of events $n_c/N$ which belong to a certain
class $c$ add up to one, because $\sum_{c=1}^{12} n_c = N$. We
therefore do not include the fraction of events without charged
leptons, $n_{0l}/N$, among our observables; note that this does not
lead to any loss of information. We thus end up with $84$ observables.

For the calculation of $\chi^2$, and also in order to improve the
performance of our artificial neural networks, we need the covariance
matrix of all $84$ observables. The variance of the total number
of events after cuts, $N$, is
\begin{equation} \label{sn}
\sigma^2(N) = N.
\end{equation}
The next twelve observables are the fractions of events $n_c/N$ that
belong to each class $c$. As mentioned before they are not
independent. The covariance between the fraction of events in two
different classes $c$ and $c'$ is then: 
\begin{equation} \label{equ:ccc}
{\rm cov} \left( \frac{n_c}{N}, \, \frac{n_{c'}}{N} \right) = \delta_{cc'}
  \frac{n_c}{N^2} - \frac{n_c \, n_{c'}}{N^3} \ \ \ (c, \, c' \in
  \{1, 2, \dots, 12\}) \,. 
\end{equation}
The covariance for identical classes ($c = c'$) equals the variance.
Note that this matrix would be singular if we included all twelve
$O_{1,c}$ among our observables. In contrast, $n_c/N$ and the total
number of events $N$ are not correlated, i.e.
\begin{equation} \label{equ:ccN}
{\rm cov} \left( \frac{n_c}{N}, \, N \right) = 0 \ \ \ (c \in
\{1,2,\dots,12\}) \,.
\end{equation}
The remaining observables can be written as averages over all events
in a given class, $O_{i,c} = \langle o_i \rangle_c$ with $o_2 =
\tau^-, \, o_3 = \tau^+, \, o_4 = b, \, o_5 = j, \, o_6 = j^2 $ or
$o_7 = H_T$. Their variances can be calculated directly from the
simulated data using the formula
\begin{equation} \label{equ:cov234567}
\sigma^2(O_{i,c}) = \frac{1}{n_c - 1} \cdot (\langle o_i^2 \rangle_c
- \langle o_i \rangle_c^2) \ \ \ (i \in \{2,3, \dots ,7\}). 
\end{equation}
Of these observables, only $\langle j \rangle_c$ and $\langle j^2 \rangle_c$
are correlated within a given class:
\begin{equation}
{\rm cov} (\langle j \rangle_c, \, \langle j^2 \rangle_c) = \frac{1}{n_c - 1}
\cdot (\langle j^3 \rangle_c - \langle j \rangle_c \, \langle
j^2 \rangle_c)\,. 
\end{equation}
Here $\langle j^3 \rangle_c$ is also determined directly from the
(simulated) events. Observables from different classes are not
statistically correlated. We also ignore the possible correlation
between $\langle \tau^- \rangle_c$ and $\langle \tau^+ \rangle_c$. The
validity of this approximation was checked for the closely related
observables $n_{c,\tau^-}/n_c$ and $n_{c,\tau^+}/n_c$ in \cite{bod4}
and should also be fine here.

\subsection{Benchmark Points}
\label{sec:Benchmarks}
\setcounter{footnote}{0} 

We look at four different reference points in the CMSSM parameter
space each yielding $O(1000)$ events after cuts for
$\unit[10]{fb^{-1}}$ of data: 
\begin{itemize}
\item[1)] $ m_0 = \unit[150]{GeV} $, \, $ m_{1/2} = \unit[700]{GeV} $,
  \, $ \tan\beta = 10 $ \, and \, $ A_0 = \unit[0]{GeV} $ 
\item[2)] $ m_0 = \unit[2000]{GeV} $, \, $ m_{1/2} = \unit[450]{GeV}
  $, \, $ \tan\beta = 10 $ \, and \, $ A_0 = \unit[0]{GeV} $ 
\item[3)] $ m_0 = \unit[1000]{GeV} $, \, $ m_{1/2} = \unit[600]{GeV}
  $, \, $ \tan\beta = 10 $ \, and \, $ A_0 = \unit[1500]{GeV} $ 
\item[4)] $ m_0 = \unit[400]{GeV} $, \, $ m_{1/2} = \unit[700]{GeV} $,
  \, $ \tan\beta = 30 $ \, and \, $ A_0 = \unit[0]{GeV} $ 
\end{itemize}
All points have a positive Higgs(ino) mass parameter $\mu$. The
resulting spectra of superparticles and Higgs bosons are given in
Table~\ref{tab:spec}.\footnote{Recall that we are not requiring the
  CMSSM to contain a Higgs boson with mass near 125 GeV in our analysis.}

\begin{table}[ht]
\begin{center}
\caption{CMSSM input parameters and selected superparticle and Higgs
  boson masses for our four benchmark points. All mass parameters are
  in GeV. Note that first and second generation sfermions with the
  same gauge quantum numbers have identical masses. Moreover,
  $m_{\tilde d_L} \simeq m_{\tilde u_L}$ in these scenarios, while
  $m_{\tilde d_R} \simeq m_{\tilde u_R}$, and $m_{\tilde \nu} \simeq
    m_{\tilde e_L}$. Similarly, $m_{\tilde \chi_1^{\pm}} \simeq m_{\tilde
      \chi_2^0}$ in all cases, while $m_{\tilde \chi_4^0} \simeq
    m_{\tilde \chi_2^{\pm}}$ are about $10$ to $30$ GeV above $m_{\tilde
      \chi_3^0}$. Finally, in all cases $m_H \simeq m_A \simeq
    m_{H^\pm}$.}
\begin{tabular}{|c||c|c|c|c|}
\hline
 & Point 1 & Point 2 & Point 3 & Point 4 \\ \hline
$m_0$ & $150$ & $2000$ & $1000$ & $400$ \\
$m_{1/2}$ & $700$ & $450$ & $600$ & $700$ \\
$A_0$ & $0$ & $0$ & $1500$ & $0$ \\
$\tan\beta$ & $10$ & $10$ & $10$ & $30$ \\
\hline
$m_{\tilde g}$ & $1570$ & $1133$ & $1407$ & $1578$ \\
$m_{\tilde u_L}$ & $1437$ & $2171$ & $1575$ & $1483$ \\
$m_{\tilde u_R}$ & $1382$ & $2161$ & $1541$ & $1430$ \\
$m_{\tilde b_1}$ & $1322$ & $1816$ & $1402$ & $1323$ \\
$m_{\tilde b_2}$ & $1371$ & $2145$ & $1529$ & $1377$ \\
$m_{\tilde t_1}$ & $1118$ & $1384$ & $1168$ & $1145$ \\
$m_{\tilde t_2}$ & $1357$ & $1824$ & $1414$ & $1366$ \\
$m_{\tilde e_R}$ & $306$ & $2005$ & $1024$ & $480$ \\
$m_{\tilde e_L}$ & $494$ & $2015$ & $1074$ & $616$ \\
$m_{\tilde \tau_1}$ & $298$ & $1988$ & $1010$ & $414$ \\
$m_{\tilde \tau_2}$ & $494$ & $2007$ & $1068$ & $606$ \\
$m_{\tilde \chi_1^0}$ & $291$ & $187$ & $249$ & $293$ \\
$m_{\tilde \chi_2^0}$ & $551$ & $344$ & $468$ & $555$ \\
$m_{\tilde \chi_3^0}$ & $848$ & $463$ & $668$ & $830$ \\
$m_h$ & $116$ & $116$ & $114$ & $117$ \\
$m_A$ & $969$ & $2040$ & $1244$ & $889$\\
\hline
\end{tabular}
\label{tab:spec}
\end{center}
\end{table}

The first parameter set has a low $m_0$, leading to small slepton
masses. The sleptons can therefore be produced on--shell in decays of
the wino--like states $\tilde \chi_2^0$ and $\tilde \chi_1^\pm$, which
in turn are produced frequently in decays of $SU(2)$ doublet squarks
$\tilde u_L, \tilde d_L$. Since these squarks can be produced either
directly or in decays of gluinos, we expect this benchmark point to
lead to relatively strong leptonic signatures. 

In contrast, the second point has $m_0^2 \gg m_{1/2}^2$, so that
squarks and sleptons have similar masses. The wino--like $\tilde
\chi_2^0$ and $\tilde \chi_1^\pm$ are still produced rather copiously
in $\tilde g$ three--body decays via virtual $SU(2)$ doublet squark
exchange; note that $\tilde g$ does not have any tree--level two--body
decay in this scenario.\footnote{Decays of the kind $\tilde g
  \rightarrow \tilde \chi_i^0 g$, which proceed via one--loop
  diagrams, are also included in SUSY-HIT \cite{bib:susyhit}, but
  their branching ratios are small.} However, $\tilde \chi_2^0$ mostly
decays into the light CP--even scalar Higgs boson $h$ plus the
lightest superparticle (LSP) $\tilde \chi_1^0$ here, while $\tilde
\chi_1^\pm \rightarrow \tilde \chi_1^0 W^\pm$ decays have a branching
ratio of $100\%$. A gluino decay on average produces $0.99$ top
(anti)quarks in this scenario, whose semileptonic decays can yield
additional hard leptons; these final states are favored since
renormalization group (RG) running and $L-R$ mixing reduce the masses
of $\tilde b_1$ and $\tilde t_1$ relative to those of the other
squarks \cite{mssm}. Nevertheless we expect the strengths of the
signals with three or more leptons to be much weaker in this scenario.
Moreover, the signal in class 6 (two leptons with opposite charge and
equal flavor) will not be enhanced relative to that in class 7 (two
leptons with opposite charge and different flavor).

The third benchmark point also has a relatively high $m_0$, so that
sleptons are again not produced in the decays of strongly interacting
superparticles. However, it also has a rather high value of $|A_0|$,
which enhances both the RG running and the $\tilde t_L - \tilde t_R$
mixing. This leads to a relatively large mass splitting between
$\tilde t_1$ and the remaining squarks. The gluino mass is chosen such
that all gluinos decay into a top and an anti--stop, or vice
versa. Since $60\%$ of all $\tilde t_1$ decay into a top quark plus
one of the neutralinos, a gluino decay therefore produces on average
$1.6$ top (anti)quarks. This yields high--multiplicity final states,
including several $b-$jets and frequently also hard leptons. However,
the $\tilde t_1$ mass is still so high that direct $\tilde t_1 \tilde
t_1^*$ pair production does not contribute appreciably to the overall
signal for supersymmetry in this scenario.

Finally, the fourth benchmark point again has $m_0 < m_{1/2}$, but
rather large $\tan\beta$. This increases the value of the $\tau$
Yukawa coupling, which in turn reduces the mass of the lighter $\tilde
\tau$ eigenstate through RG effects as well as $\tilde \tau_L - \tilde
\tau_R$ mixing. As a result, $\tilde \tau_1$ is significantly lighter
than the other sleptons in this scenario. Although $m_{\tilde e_R} <
m_{\tilde \chi_2^0} \simeq m_{\tilde \chi_1^{\pm}}$, essentially no
$\tilde \chi_1^\pm$ and just $0.2\%$ of all $\tilde \chi_2^0$ decay
into a first or second generation charged slepton. The reason is that
$\tilde \chi_1^\pm$ and $\tilde \chi_2^0$ are both wino--like, whereas
$\tilde e_R$ and $\tilde \mu_R$ are $SU(2)$ singlets. On the other
hand, $\tilde \tau_1$ has a significant $SU(2)$ doublet component. As
a result, about $77\%$ each of all $\tilde \chi_1^\pm$ and $\tilde
\chi_2^0$ decay into a $\tilde \tau_1$, which in turn always decays
into a $\tau$ lepton and an LSP $\tilde \chi_1^0$; the remaining
$\tilde \chi_2^0$ decay mostly into $h+\tilde \chi_1^0$, whereas the
remaining $\tilde \chi_1^\pm$ decay into $W^\pm + \tilde
\chi_1^0$. The decay products of strongly interacting superparticles
therefore frequently contain one or more $\tau$ lepton(s). On the
other hand, the branching ratios for gluino decays into third
generation squarks are again enhanced, so that each gluino decay
produces on average $0.99$ top (anti)quarks. In this scenario
supersymmetric events therefore can contain relatively soft leptons
from leptonic $\tau$ decays and/or hard leptons from semileptonic top
decays, in addition to $\tau-$ and/or $b-$jets.

\begin{table}[!h]
\begin{center}
  \caption{Distribution of class events in percent into the twelve
    mutually exclusive lepton classes (determined from
    $\unit[500]{fb^{-1}}$ of simulated data), and their statistical errors 
    for an integrated luminosity of $\unit[10]{fb^{-1}}$, for the four
    benchmark points.} 
\begin{tabular}{|l||c|c|c|c|}
\hline
Class & Point 1 & Point 2 & Point 3 & Point 4 \\ \hline
1. $0l$ & $46.6 \pm 2.1$ & $47.1 \pm 2.1$ & $54.3 \pm 2.2$ & $61.2 \pm 2.5$ \\
2. $1l^-$ & $12.2 \pm 1.1$ & $16.2 \pm 1.2$ & $15.0 \pm 1.1$ & $12.4
\pm 1.1$ \\
3. $1l^+$ & $18.0 \pm 1.3$ & $16.9 \pm 1.3$ & $16.8 \pm 1.2$ & $15.5
\pm 1.3$ \\
4. $2l^-$ & $1.6 \pm 0.4$ & $2.8 \pm 0.5$ & $1.7 \pm 0.4$ & $1.3 \pm
0.4$ \\
5. $2l^+$ & $3.9 \pm 0.6$ & $3.3 \pm 0.6$ & $2.2 \pm 0.4$ & $1.9 \pm
0.4$ \\
6. $l^+_i l^-_i$ & $7.7 \pm 0.8$ & $4.3 \pm 0.6$ & $3.2 \pm 0.5$ &
$2.8 \pm 0.5$ \\
7. $l^+_i l^-_{j;\ j \neq i}$ & $3.5 \pm 0.6$ & $5.0 \pm 0.7$ & $3.8
\pm 0.6$ & $2.9 \pm 0.5$ \\
8. $l^-_i l^-_j l^+_j$ & $1.9 \pm 0.4$ & $1.4 \pm 0.4$ & $0.9 \pm 0.28$
& $0.6 \pm 0.25$ \\
9. $l^+_i l^+_j l^-_j$ & $3.4 \pm 0.6$ & $1.4 \pm 0.4$ & $1.0 \pm 0.3$
& $0.7 \pm 0.27$ \\
10. $l^-_i l^-_j l^{\pm}_{k;\ k \neq j, i \ {\rm for} \ +}$ & $0.1 \pm
0.10$ & $0.5 \pm 0.22$ & $0.3 \pm 0.16$ & $0.2 \pm 0.14$ \\
11. $l^+_i l^+_j l^{\pm}_{k;\ k \neq j, i \ {\rm for} \ -}$ & $0.2 \pm
0.14$ & $0.5 \pm 0.22$ & $0.3 \pm 0.16$ & $0.3 \pm 0.17$ \\
12. $4l$ & $0.8 \pm 0.27$ & $0.6 \pm 0.24$ & $0.5 \pm 0.21$ & $0.2 \pm
0.14$ \\
\hline
\end{tabular}
\label{tab:classEvents}
\end{center}
\end{table}

Our qualitative expectations are confirmed by
Table~\ref{tab:classEvents}, which lists the fractions of events (in
percent) that are assigned to the twelve event classes. We also note
that all benchmark points predict that more positively than negatively
charged leptons are produced. This results because the proton contains
more $u-$quarks than $d-$quarks, so that the production of $\tilde
u-$squarks is enhanced relative to that of $\tilde d-$squarks. The
asymmetry between positively and negatively charged leptons becomes
large when squark production contributes prominently to the total
event sample (after cuts) and (some) first generation squarks have
sizable semi--leptonic decay branching ratios. Both conditions are
satisfied for Point 1 and, to a lesser extent, in Point 4, whereas
Point 2 has a very small asymmetry owing to the large squark masses
and small leptonic branching ratios of $\tilde \chi_1^\pm$. 

Since we assume exact conservation of $R-$parity, the LSP is
stable. In the four benchmark scenarios, it is the lightest neutralino
$\tilde \chi_1^0$. Each supersymmetric event will therefore contain at
least two $\tilde \chi_1^0$, which are not detected, leading to the
celebrated missing transverse momentum signature. Indeed, our cuts
always include the requirement of a sizable missing $p_T$, the exact
value of the cut depending on the number of charged leptons in the
event; we again refer to the Appendix of ref.\cite{bod4} for further
details.\footnote{Note that we apply a ``$Z$ veto'' cut on events of
  class 6, which is not applied on events in class 7. Without this
  cut, in a theory respecting $e-\mu$ universality, as the CMSSM does,
  class 6 would have to contain at least as many events as class 7,
  since uncorrelated lepton pairs would contribute equally to both
  classes, while correlated same flavor lepton pairs only contribute
  to class 6.}

We find the following numbers of events for an integrated luminosity
of $\unit[10]{fb^{-1}}$:
\begin{itemize}
\item[1)] 1940 events before and 1080 after cuts
\item[2)] 4080 events before and 1047 after cuts
\item[3)] 1970 events before and 1135 after cuts
\item[4)] 1618 events before and 991 after cuts
\end{itemize}
These numbers have been obtained from a simulation corresponding to an
integrated luminosity of $\unit[500]{fb^{-1}}$.

Evidently the number of events after cuts differs much less between
the four benchmark points than the raw event number does. In
particular, for the second benchmark point the large number of events
before cuts is due to the small masses of the electroweak charginos
and neutralinos. In principle these final states -- in particular,
$\tilde \chi_2^0 \tilde \chi_1^\pm$ production -- can lead to final
states with three charged leptons and little hadronic activity
\cite{bib:3l}, which could contribute to our event classes 8 or
9. Recall, however, that $\tilde \chi_2^0$ has a very small leptonic
branching ratio in this scenario. The $\tilde \chi$ pair events therefore
have a very low cut efficiency, i.e.~most of these events do not pass
our cuts. 

It is amusing to note that the total number of events after cuts alone
would evidently not be sufficient to distinguish between benchmark
points 1 and 3, nor between points 2 and 4. This illustrates the need
to include (many) more observables when trying to discriminate
scenarios, let alone for the quantitative determination of the values
of the free parameters. Indeed, Table~\ref{tab:classEvents} shows that
the event fractions in the twelve classes are quite sufficient to
distinguish between the four benchmark scenarios. However, we do not
merely wish to distinguish between benchmark points that are well
separated in parameter space; we wish to quantitatively determine the
values of the underlying parameters. This is discussed in the
subsequent Section.

\section{Strategies for Determining the Parameters}
\label{sec:Strategies}
\setcounter{footnote}{0}

In this Section we discuss the strategies we employed to extract the
values of the four CMSSM parameters from our 84 observables for the
four benchmark points described in the previous Subsection. We first
describe the construction of an artificial neural network (ANN), and
then an attempt based on the minimization of an overall $\chi^2$
variable.

\subsection{Neural Network}
\label{sec:NeuralNetwork}

Artificial neural networks have been used in high energy physics since
more than 25 years \cite{oldreview}. Originally they were designed for
comparatively simple tasks, like reconstructing single tracks from hit
patterns in wire chambers. By now ANNs are used for a wide variety of
purposes, from optimizing experimental searches for superparticles
\cite{recent_nn} to parameterizing parton distribution functions
\cite{nnpdf}. However, we are not aware of a previous application of
ANNs to the LHC inverse problem.

Here we describe how to construct artificial neural networks that can
find relations mapping the measured observables onto the CMSSM
parameters whose values we wish to determine. Mathematically speaking,
an ANN is a function mapping input parameters, to be provided by the
user, onto output parameters, whose values the user wishes to
determine. It can learn this function from training sets via a
well--defined algorithm. A training set consists of input values (the
observables of simulated experiments) and the corresponding output
values (the CMSSM parameters). Note that the parameters that are
the input into an event generation program (in our example, the CMSSM
parameters) are the output of the neural network, whereas the output of
the event simulation (our 84 observables) are input of the neural
network. It is hoped that by training the neural network on
sufficiently many sets of input and output parameters, it will learn
to derive the correct values of the output parameters also for sets of
input parameters that are not contained in the training set. We will
show that, at least for our four benchmark scenarios, this indeed
works quite well.

As noted above, a neural network is a mathematical function;
``learning'' means that the numerical coefficients defining this
function are adjusted. By choosing different values for these
coefficients, in principle nearly every possible function can be
reproduced with some accuracy. During the learning process these
coefficients are set such that the deviations between the calculated
network outputs for the specific training inputs and the known desired
(correct) outputs are reduced as far as possible. If the training set
is a fair representation of the investigated CMSSM parameter space, at
the end the resulting neural network function should be able to
compute the correct CMSSM parameters for an arbitrary set of input
parameters, assuming that the latter indeed can be reproduced in the
CMSSM.

\subsubsection{Set--up}
\label{sec:setup}
\setcounter{footnote}{0}

\begin{figure*}[h]
\centering
\resizebox{0.65\textwidth}{!}{
  \includegraphics{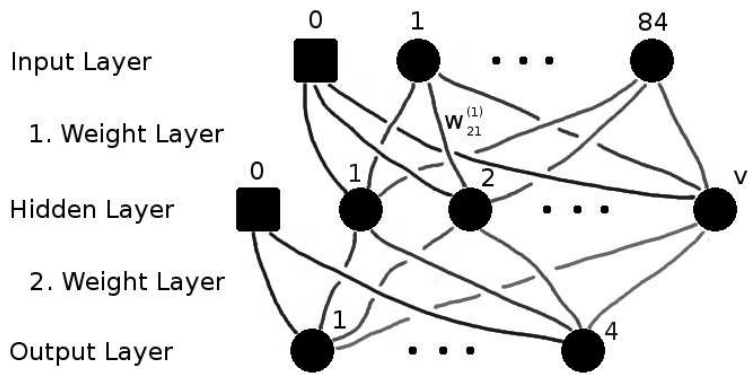}}
\caption{Neural network with one layer of hidden ``neurons'', i.e.~two
  weight layers. The squares represent the neurons with constant
  output and the circles the neurons with variable output, which
  depends on the network input. All neurons of adjacent layers are
  connected to each other. Each connection can have a different
  weight, e.g.~the connection between the first input neuron and the
  second hidden neuron has the weight $w_{21}^{(1)}$.}
\label{fig:NeuralNetwork}
\end{figure*}

General information about neural networks can e.g.~be found in
\cite{bib:BookNeuralNetwork}. Generally speaking, an ANN consists of a
set of ``neurons'', collected in ``layers'', and ``weights'' assigned
to the connections between these neurons. A neuron can be described by
a simple function computing one output value from one input value. At
least two layers of neurons are needed: one for the input, and one for
the output. In the input layer, there is one neuron for each
observable, plus a ``constant'' neuron which always outputs 1; this is
needed in order to be able to describe a constant function, i.e.~a
component of the network's output that is independent of the
input. Similarly, there is one output neuron for each quantity that we
wish to determine. The input of the neurons in the input layer are
normalized versions of our observables, and the output of the neurons
in the output layer gives the network's prediction for a normalized
version of the quantity whose value we wish to determine; this
normalization is described below. In between the input and output
layers, there can be an arbitrary number of layers of hidden neurons;
in physics applications this number is usually one or two. All neurons
in adjacent layers are connected to each other, and a numerical weight
is assigned to each of these connections. These weights are the
coefficients mentioned above, whose values are to be determined during
the training of the network.\footnote{The term ``artificial neural
  network'' reflects the fact that this construction yields a very
  simple model of actual biological neural networks; however, the
  (dis)similarity between ANNs and actual brains is immaterial for our
  discussion.}

The ANN we construct here contains a single layer of hidden neurons,
as shown in Figure \ref{fig:NeuralNetwork}. For all neurons in the
input and output layers, the output is simply equal to the input. In
our case the input layer consists of 85 neurons, one for each
(normalized) observable with input values $x_i$, $i = 1, \, \dots, \,
84$; the output of the $i-$th neuron in this layer is therefore also
given by $x_i$. As noted above, there is an additional input neuron
with a fixed output, $x_0 = 1$.

The second layer consists of a yet to be determined number $v$ of
``hidden'' neurons with input values $z_1, \, \dots, \, z_v$. These
input values are weighted sums of the output of all neurons in the
input layer:
\begin{equation} \label{equ:hiddenNeuronInput}
z_a = \sum\limits_{i = 1}^{84} w_{ai}^{(1)} x_i + w_{a0}^{(1)} =
\sum\limits_{i = 0}^{84} w_{ai}^{(1)} x_i \, . 
\end{equation}
Here and in the following we will use $i, j, \dots$ to label the input
neurons, $a, b, \dots$ to label the hidden neurons, and $r, s, \dots$
for the output neurons. Each hidden neuron processes its input by
applying the function $\tanh(z_a)$. This function is commonly chosen
for the hidden neurons, because it exhibits approximately linear
behavior for small $|z_a|$ but saturates at $\pm 1$ for large values
of $|z_a|$. This ensures that the ANN can reproduce a wide variety of
functions.\footnote{Due to the saturation of the $\tanh$ function at
  large absolute values of its arguments, an ANN constructed in this
  way may have difficulty reproducing a function which grows or
  shrinks very quickly when its input variables are varied. Recall,
  however, that the input of the ANN (during training) is the output
  of event simulation. In extensions of the Standard Model, the number
  of events of a given type expected at the LHC typically changes
  quickly when the model parameters are varied. This implies that the
  true values of the model parameters change relatively slowly when
  the observables are varied. Our ANN should be well suited to
  describe this kind of behavior.}

The last layer of the ANN consists of the output neurons, four in
number for CMSSM with values $y_1, \, \dots, \, y_4$.\footnote{For
  reasons that will be explained below, at the end we actually do not
  use one common network for all four CMSSM parameters but instead
  four separate ANNs with one output parameter each.} Their input
values are calculated by weighted sums over the output of the hidden
neurons:
\begin{equation} \label{equ:outputNeuronInput}
y_r = \sum\limits_{a = 1}^v w_{ra}^{(2)} \tanh(z_a) + w_{r0}^{(2)} =
\sum\limits_{a = 0}^v w_{ra}^{(2)} \tanh(z_a) \,;
\end{equation}
the output of the zeroth hidden neuron has been fixed to $\tanh(z_0) =
1$. These four output values should reproduce the (properly
normalized) values of the CMSSM parameters; the normalization is
explained below.

Two ``weight layers'' connect the three layers of neurons. The first
weight layer connects each input neuron with each hidden neuron
(except for the zeroth hidden neuron whose output value is fixed to
$1$). The weights $w_{ai}^{(1)}$ with $a = 1, \, \dots, \, v$ and $i =
0, \, \dots,\, 84$ are assigned to these connections. Similarly, the
connections between the hidden and output neurons have weights
$w_{ra}^{(2)}$ with $ r = 1, \, \dots, \, 4 $ and $ a = 0, \, \dots,
\, v $ and form the second weight layer. Evidently the number of
hidden neurons determines the number of free parameters describing the
ANN. The goal of the training process is to find appropriate (ideally,
optimal) values for these weights. Before we describe the algorithm
used to set the weights, we discuss the normalization of the input and
output values, and the initialization of the weights.

\subsubsection{Normalization}
\label{sec:Normalization}

The 84 inputs into our ANN are obtained by normalizing the 84 measured
observables. This simplifies the initialization of the weights (see
below), which in turn results in a shorter learning process. To that
end, both the input and output values of the ANN should be
(dimensionless) ${\cal O}(1)$ quantities. Because the hyperbolic
tangent function also naturally leads to hidden layer output values of
${\cal O}(1)$, the weights in both layers could be of a similar
order. In most cases the $\tanh$ functions will then not be in the
saturation regime, where it becomes almost independent of the input
value. In general this should speed up the learning process.

We normalize the 84 input quantities $O_i$ independently. Let $O_i^n$
be the value of observable $O_i$ for the $n-$th training set, with $n
= 1, \, \dots, \, N$. The mean value $\overline{O}^{tg}_i$ and the
variance $(\sigma_i^{tg})^2$ over all $N$ training sets are then:
\begin{equation} \label{equ:ave}
\overline{O}^{tg}_i = \frac{1}{N} \sum\limits_{n = 1}^N O_i^n\,;
\end{equation}
\begin{equation} \label{equ:VarianceInputNormalization}
(\sigma_i^{tg})^2 = \frac{1}{N - 1} \sum\limits_{n = 1}^N (O_i^n -
\overline{O}^{tg}_i)^2 \,.
\end{equation}
From these, we compute normalized inputs $x_i$ for the ANN:
\begin{equation} \label{equ:InputNormalization}
x_i = \frac{O_i - \overline{O}^{tg}_i}{\sigma_i^{tg}}\,.
\end{equation}
The mean value of the $x_i$ over the training sets is thus zero, with
a standard deviation equal to one; hence the (absolute values of) the
inputs should be ${\cal O}(1)$, as desired.\footnote{Equivalently, one
  could use the original observables $O_i$ as input, and assign the
  function (\ref{equ:InputNormalization}) to the $i$--th input neuron
  instead of the identity function.}

Since we want the ANN outputs $y_r$ to also be ${\cal O}(1)$ in
absolute value, they are related to the CMSSM parameters $r \in \{m_0,
\, m_{1/2}, \, \tan\beta, \, A_0\}$ which we wish to determine via
\begin{equation} \label{equ:OutputNormalization}
y_r = \frac{r - \overline{r}^{tg}}{\sigma_r^{tg}}\,.
\end{equation}
The averages $\overline{r}^{tg}$ and the variances $(\sigma_r^{tg})^2$ are
calculated analogously to eqs.(\ref{equ:ave}) and
(\ref{equ:VarianceInputNormalization}), respectively.

The normalization of both the input and the output is thus calculated
from the training sets. The same normalization is then used for the
control sets, and for every other ANN input including the (actual or
simulated) measurement. The implicit assumption is that the training
sets are a good representation of the investigated CMSSM parameter
space. If this condition is not satisfied, the ANN results for the
CMSSM parameters are in any case likely to be quite inaccurate.

\subsubsection{Initialization}
\label{sec:Initialization}

The normalization of the overall input and output of the ANN ensures
that these quantities are ${\cal O}(1)$ in absolute size, without
preference for either positive or negative values. The weights in the
two weight layers should be initialized such that these conditions are
also satisfied for the inputs of the neurons in the hidden and output
layers. One simple, and by experience quite effective, way to ensure
this is to initialize all weights with Gaussian random numbers with
mean zero and variance $\sigma^2$ equal to the number of neurons in
the layer beneath the given weight layer. Hence we set the Gaussian
variance for the weights in the first weight layer to $1/85$, and for
the second layer to $1/(v + 1)$, where $v$ is as before the number of
hidden neurons.

\subsubsection{Learning Procedure}
\label{sec:Learning}
\setcounter{footnote}{0}

The initial ANN will in general provide a very poor approximation of
the desired function. In the case at hand, the initial ANN will most
likely give values for the CMSSM parameters that are very far from the
true values.

The ANN therefore needs to be ``trained'', so that it can ``learn''
to closely reproduce the desired function. To that end it has to be
confronted with sufficiently many ``training sets'', where {\em both}
the input and the desired output are known. In the case at hand, this
means that we had to simulate many sets of CMSSM parameters, and
compute the corresponding values of our 84 observables, along the
lines described in Sec.~2. More details on the choice of the training
sets are given below.

An ANN which is trained with specific training sets reproduces the
desired output for these training sets more and more exactly with
every learning step. If the network includes sufficiently many hidden
neurons it will eventually simply ``memorize'' the training sets,
i.e.~reproduce their outputs exactly, if the training runs long
enough. This may seem desirable at first sight, but actually it is
not. The task of the ANN is to interpolate between the training sets,
i.e.~it should provide (approximately) the correct output also for
inputs that are {\em not} part of the training sets. Experience shows
that at some point further improvement in the reproduction of the
training sets {\em degrades} the performance of the ANN when applied
to different inputs.

In order to determine when the training of the ANN should be
terminated one therefore also needs ``control sets''. These are
generated exactly like the training sets, but they are {\em not} used
in the training of the ANN. Instead, they are used to define a
``control error''. The training of the ANN is stopped when this
control error reaches its minimum. This strategy ensures good
performance of the neural network as long as both the training and the
control sets are good representations of the investigated CMSSM
parameter space.

We use the following normalized control error:
\begin{equation} \label{equ:NormalizedError}
\tilde{F} = \sqrt{\frac{\sum\limits_{n = 1}^{M} \left( \vec{y}^n -
      \vec{k}^n \right)^2}{\sum\limits_{n = 1}^{M} \left(
      \overline{\vec{k}} - \vec{k}^n \right)^2}}\,, 
\end{equation}
where $M$ is the number of control sets, $\vec{y}^n$ are the four
output values combined in one vector, calculated by the ANN from the
input values of the $n$--th control set, $\vec{k}^n$ are the correct
output values and $\overline{\vec{k}}$ is the mean value over the
output values of all control sets:
\begin{equation}
\overline{\vec{k}} = \frac{1}{M} \, \sum\limits_{n = 1}^{M} \vec{k}^n.
\end{equation}
The only quantities in eq.(\ref{equ:NormalizedError}) that change in
the course of the training are the vectors $\vec{y}^n$, which depend
on the current values of the weights defining the ANN. These weights
are modified iteratively, such that an error analogous to that defined
in eq.(\ref{equ:NormalizedError}), but computed from the training sets
rather than from the control sets, is minimized. Mathematically this
amounts to minimizing a (complicated) function of the (many) variables
$w_{ai}^{(1)}, \, w_{ra}^{(2)}$. We do this using the ``conjugate
gradient'' algorithm, as described in Appendix A.

Recall that $\vec{k}^n$ and $\vec{y}^n$ are normalized output
values. If the ANN has several outputs, the contribution of each
output variable to the total error therefore depends on the spread of
this variable within the training sets. The algorithm for adjusting
the weights ensures that the total training error decreases during the
training; however, this does not guarantee that the performance for
each individual output, i.e.~for each component of $\vec{y}$, improves
monotonically. We therefore found it convenient to define four
separate ANNs for the four CMSSM parameters, as already noted. For
each of our ANNs the vectors $\vec{y}$ and $\vec{k}$ collapse to
simple real numbers.  The splitting into four separate network also
offers the possibility of a further specialization of each
network. Furthermore we have smaller networks, whose training take
less time than the training of one big network with a higher number of
hidden neurons. Note, however, that all four ANNs have identical
inputs, i.e.~each of them has 84 input neurons.

The training of the ANNs thus proceeds as follows. The initial values
of the weights are used to define initial errors for the training and
control sets. The weights are then adjusted iteratively, such that the
error computed for the training sets is minimized. After each learning
step the control error is calculated. The learning process is stopped
when this control error reaches its minimum.\footnote{Had we used a
  single ANN, with a combined control error as defined in
  eq.(\ref{equ:NormalizedError}), the training would have to be
  stopped when the {\em total} error is minimal. This does not
  guarantee that each of the four individual errors, on the four CMSSM
  parameters, is minimal. This is another argument in favor of using
  four separate ANNs.} Since the control error is not used for the
determination of the new weights, it need not decrease monotonically
during the iteration. The learning procedure should therefore not be
stopped until one can be reasonably sure that the {\em absolute}
minimum of the control error has been passed. Finally, the weights
have to be re--set to the values that gave this absolute minimum.

\begin{figure*}[h]
\centering
\resizebox{0.5\textwidth}{!}{
  \includegraphics{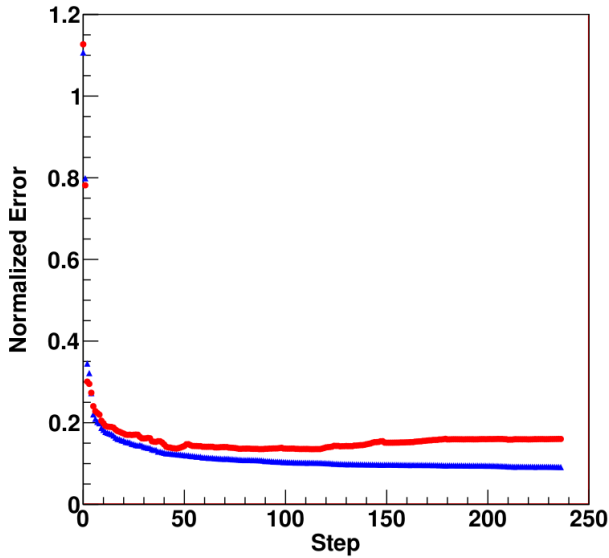}
}
\caption{Error evolution of a neural network with $20$ hidden neurons
  for the parameter $m_{1/2}$ of benchmark point 3. The blue triangles
  show the normalized training error and the red dots the
  corresponding normalized control error. The minimum of the control
  error is reached with $0.136$ at learning step $107$.} 
\label{fig:Fehlerentwicklung}
\end{figure*}

This is illustrated in Fig.~\ref{fig:Fehlerentwicklung}. The triangles
(blue) show the normalized training errors and the dots (red) the
control errors. We see that both errors initially decrease very
quickly. The control error reaches a first minimum after about $45$
learning steps, increases again, and finally reaches its absolute
minimum after $107$ steps. In contrast, the training error keeps
decreasing until the iteration is stopped after $235$ steps; the
control error at the end of the iteration is nearly $10\%$ larger than
its absolute minimum. At the end the ANN is therefore re--set to its
status after $107$ learning steps.

\subsubsection{Training and Control Sets}
\label{sec:Sets}

The performance of the trained ANNs obviously depends on the quality
of the training and control sets. Note first of all that these sets
are supposed to show the true relation between the 84 observables and
the CMSSM parameters. The errors on the observables in the training
and control sets, in particular the error due to finite Monte Carlo
statistics, should therefore be (much) smaller than the statistical
accuracy of the (simulated) measurement that the ANNs are finally
meant to analyze. We therefore used $\unit[500]{fb^{-1}}$ of simulated
data for the training and control sets; of course, the number of
events is then rescaled to the assumed luminosity of the
``measurement''.

Moreover, we already emphasized that the training and control sets
should be good representations of the investigated CMSSM parameter
space. Therefore for each benchmark point we set different ranges for
the input parameters of the simulation; the actual CMSSM parameters
for each training and control set are chosen randomly using a flat
distribution.\footnote{Furthermore each simulation in Herwig++ is done
  with a different flatly, randomly chosen seed to prevent any
  possible correlations.}

In general it should be easier to determine the parameters $m_0$ and
$m_{1/2}$, since they influence the superparticle spectrum more
strongly than the other two continuous parameters, $A_0$ and
$\tan\beta$, do. One combination of $m_0$ and $m_{1/2}$ can e.g.~be
determined with some accuracy simply from the total rate of signal
events after cuts. Next, one could use some inclusive kinematical
observable (e.g.~$H_T$), and/or the jet multiplicity in signal events,
to determine the region of interest in the $(m_0, m_{1/2})$ plane.  In
the following we therefore use smaller parameter ranges for these
parameters than for $A_0$ and $\tan\beta$. We do this just to save CPU
hours. This slight ``cheat'' allows us to generate, with our limited
computer resources, sufficiently many training and control sets in the
vicinity of our benchmark points to allow good interpolation. The
above qualitative discussion indicates that our ANNs could most likely
also be trained on the ``entire'' CMSSM parameter space. However, we
would then need to generate (many) more training and control
sets. Moreover, the ANNs would presumably need more hidden neurons,
making their training even more time consuming.

We use the following parameter ranges for our four benchmark points:
\begin{itemize}
\item[1)] $ m_0: \, \unit[100 - 350]{GeV} $, \, $ m_{1/2}: \,
  \unit[660 - 740]{GeV} $, \, $ \tan\beta: 5 - 45 $ \, and \, $ |A_0|
  \leq 2 \cdot m_0 $ 
\item[2)] $ m_0: \, \unit[1850 - 2200]{GeV} $, \, $ m_{1/2}: \,
  \unit[410 - 490]{GeV} $, \, $ \tan\beta: 5 - 45 $ \, and \, $ |A_0|
  \leq 2 \cdot m_0 $ 
\item[3)] $ m_0: \, \unit[850 - 1200]{GeV} $, \, $ m_{1/2}: \,
  \unit[560 - 640]{GeV} $, \, $ \tan\beta: 5 - 45 $ \, and \, $ |A_0|
  \leq 2 \cdot m_0 $ 
\item[4)] $ m_0: \, \unit[300 - 550]{GeV} $, \, $ m_{1/2}: \,
  \unit[660 - 740]{GeV} $, \, $ \tan\beta: 10 - 50 $ \, and \, $ |A_0|
  \leq 2 \cdot m_0 $ 
\end{itemize}
The range for $A_0$ depends on the value chosen for $m_0$ in order to
make sure that there is no problem with the generation of the
superparticle spectrum in SOFTSUSY.\footnote{$|A_0| \gg m_0$ could
  e.g.~result in tachyonic sfermions.} We simulate slightly more than
$1,000$ training sets and around $300$ control sets for each benchmark
point. We checked that adding additional control sets does not improve
the results; recall that these sets are only needed to determine when
the training of the ANNs should be terminated. The total number of
simulated parameter sets was determined by our available computing
resources. Note, however, that the average distance between
neighboring values of any CMSSM parameter in the training sets is much
smaller than our final estimate of the error with which this parameter
can be determined by our ANNs (see Sec.~4), at least for the smaller
integrated luminosity of $\unit[10]{fb^{-1}}$; i.e.~the final
(multi--dimensional) error ellipsoid should already contain many
training sets. Further increasing the number of training sets is
therefore not likely to significantly improve the final performance of
our ANNs.

\subsubsection{Improving the Performance}
\label{sec:Improvement}
\setcounter{footnote}{0}

The ``measured'' observables resulting from each simulated parameter
set have statistical uncertainties, which are saved in the covariance
matrix. Recall, however, that only the values of the observables
themselves, but not the corresponding covariance matrix, is used as
input into our ANNs. It would be desirable if the ANN knew with what
precision an input observable is determined in order to be able to
decide how important its value is for the determination of the CMSSM
parameters. Recall in particular that different observables can have
quite different (relative) errors; see
Table~\ref{tab:classEvents}. All else being equal, observables with
larger uncertainties should contribute with smaller weights. It is
therefore quite evident that knowledge of the covariance matrix should
improve the performance of our ANNs significantly.

Recall that we use $\unit[500]{fb^{-1}}$ of simulated data for the
training and control sets. This means that the Monte Carlo (theory)
error is essentially negligible when comparing with simulated
measurements based on an integrated luminosity of
$\unit[10]{fb^{-1}}$. We will also show results for simulated
measurements which also assume an integrated luminosity of
$\unit[500]{fb^{-1}}$. Here the Monte Carlo theory error is still
significant; our computer resources do not allow us to reduce it
further.

Directly feeding the (non--vanishing) entries of the covariance matrix
as additional inputs into our ANNs would greatly increase their
complexity. Instead, we take two measures in order to include the effect
of (statistical) uncertainties.

First, we simply omit very noisy observables, i.e.~quantities that
have very large errors. To that end, as in the $\chi^2$ calculation in
\cite{bod4}, we require a minimal number of events of a given class
$c$ for taking the observables $O_{i,c}$ into account. Obviously
observables computed from a small number of events will have a large
error. For the $\unit[500]{fb^{-1}}$ training and control sets the
observables $O_{i,c}$ are only included if the total number of events
in a given class after cuts satisfies $n_c \geq 500$; the total number
of events is only included if it exceeds 50 (in practice this is
always the case). For the $\unit[10]{fb^{-1}}$ ``measurements'' the
corresponding thresholds are changed to $10$ and $1$,
respectively. The luminosity dependence of these thresholds means that
for a specific set of CMSSM parameters the same observables are
included (up to statistical fluctuations).

Recall that an ANN is defined with a fixed number of input neurons,
therefore we still have to assign some value to each observable. The
normalized input value of an observable which does not fulfill the
minimal number of events is therefore set to zero. This means that
this input neuron does not contribute to the weighted sums which are
calculated within the neural network, independent of the weight of the
connections exiting this neuron.

This procedure ensures that all non--zero inputs into our ANNs should
have some statistical power for determining the CMSSM parameters we
are after. However, this still does not tell the ANNs the relative
accuracies of these observables. To that end, we use our knowledge of
the covariance matrix of the observables for a particular training set
to determine an $84$ dimensional (correlated) Gaussian distribution.
Here both the mean values and the covariance matrix are taken from the
training sets, independent of the luminosity of the (simulated)
measurement to which the ANN will eventually be applied. Of course,
the overall width of these multi--dimensional Gaussian distributions
(one for each original set of CMSSM parameters chosen for a given
training set) should scale like the inverse square root of the
integrated luminosity. However, the purpose of this trick is to teach
the ANN the {\em relative} size of the errors on the various
observables that are fed into the ANN, so that it can assign
appropriate weights. This {\em relative} weighting should be
independent of the luminosity.

For each combination of CMSSM parameters chosen for the training sets,
we then randomly generate $100$ sets of observables from the
corresponding up to $84$ dimensional Gaussian
distribution.\footnote{Note that these ``satellite sets'' are produced
  directly from the multi--dimensional Gaussian; no additional event
  generation is needed here.} From the point of view of the ANNs,
these sets have (slightly) different inputs, but exactly {\em the
  same} outputs (CMSSM parameters). Altogether each ANN is thus
trained on slightly more than $100,000$ sets of inputs yielding slightly
more than $1,000$ different outputs.

It is interesting to note that this enlarged set of training sets can
no longer be described as a function in the mathematical sense: due to
statistical fluctuations, there might be training sets with the same
(sets of) observable(s) but {\em different} output values.\footnote{Of
  course, the ANNs still {\em are} functions, assigning a unique
  output value to each set of input values.} In fact, for each
observable we have a certain range of values, depending on the
corresponding variance, that can lead to the same CMSSM
parameters. The bigger this range is the less important the input
value should be for the parameter determination. If the variance is
large, during the learning process the ANNs are confronted with
strongly varying input values for the same output value. This should
allow them to recognize that this observable cannot contribute much to
the determination of the CMSSM parameters, by reducing the appropriate
weight. So the creation of Gaussian distributed variants of the
original training set should lead to a higher weighting of input
values with small errors.

\subsubsection{Output Error}
\label{sec:Error}

We finally end up with four trained ANNs per benchmark point, one for
each CMSSM parameter. The 84 observables obtained from any (simulated)
experiment can now be used as input for the ANNs, which will then
produce their estimates of the corresponding CMSSM parameters. In
order to be able to judge the accuracy of these estimates, it is
crucial to also obtain estimates for the uncertainties; ideally we
would like to be able to quote the parameters with well--defined
(Gaussian) statistical errors. Our ANNs by themselves do not provide
such estimated errors.

\begin{figure*}[h!]
\centering
\resizebox{0.5\textwidth}{!}{
\includegraphics{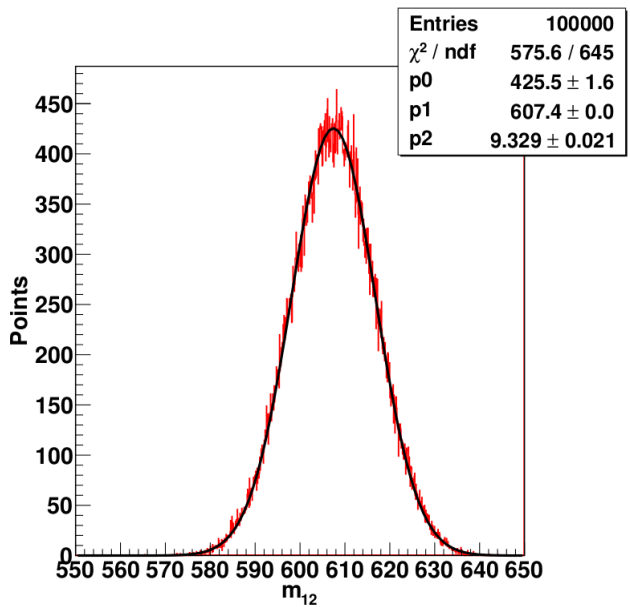}
}
\caption{One--dimensional distribution of reconstructed $m_{1/2}$
  values formed by feeding an ANN with $100,000$ Gaussian distributed
  versions of $\unit[10]{fb^{-1}}$ measurements for benchmark point
  3. The fitted Gaussian distribution has the form $g(m_{1/2}) = p_0
  \cdot \exp(-1/2 [(m_{1/2} - p_1)/p_2]^2)$, i.e.~$p_1$ is the mean
  value and $p_2$ is the standard deviation. The neural network
  contains 20 hidden neurons and underwent 107 learning steps with a
  final control error of 0.136. The true value is $m_{1/2} =
  \unit[600]{GeV}$.}
\label{fig:Gaussian1}
\end{figure*}

We investigated two different ways to determine the errors on the
outputs. The first possibility is to expand the set of input values of
the measurement to multiple Gaussian distributed input sets. As in the
construction of the ``satellite sets'' from the original training
sets, the known covariance matrix of the measurement can be used to
create a multitude of Gaussian distributed versions of the (simulated)
measurement. Each version is then used as an input for the ANNs. If
Gaussian statistics is applicable, the ANNs' outputs should also form
Gaussian distributions. The square root of the variance of an output
distribution then gives the error of the corresponding CMSSM
parameter.

This is illustrated in Fig.~\ref{fig:Gaussian1}, which shows the
distribution of reconstructed $m_{1/2}$ values for benchmark point
3. This distribution has been obtained by feeding the appropriate ANN
with $100,000$ Gaussian distributed versions of the simulated
$\unit[10]{fb^{-1}}$ measurement. We see that the true value $m_{1/2}
= \unit[600]{GeV}$ is recovered without significant off--set.
Moreover, the distribution can be described very well through a
Gaussian fit. The result of this ANN for benchmark point 3 is thus
$m_{1/2} = \unit[(607.4 \pm 9.3)]{GeV}$.

In order to determine the covariance matrix between the four CMSSM
parameters we fit two--dimensional Gaussians to the distribution of
each pair of outputs. To that end we feed the same sets of input
values into both relevant ANNs; each set of input values then gives
one pair of output values. We repeat that for all Gaussian distributed
input sets. The set of output pairs then forms a two--dimensional
(correlated) distribution and the corresponding variances and the
covariance can be determined through a fit. Within rounding errors,
e.g.~due to different binning, the variances should be exactly the
same as the ones determined from one--dimensional Gaussians.

\begin{figure*}[h!]
\centering
\resizebox{1.\textwidth}{!}{
\includegraphics{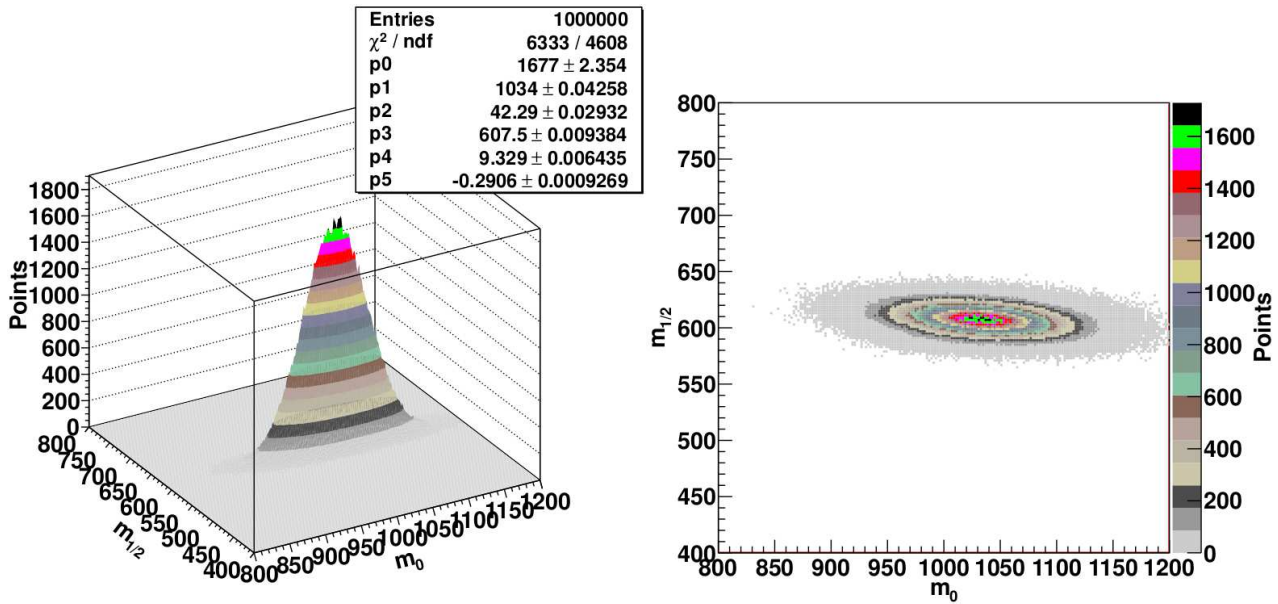}
}
\caption{Two--dimensional distribution of reconstructed $(m_0,
  m_{1/2})$ pairs (in GeV) formed by feeding both relevant ANNs with
  $1,000,000$ Gaussian distributed versions of $\unit[10]{fb^{-1}}$
  measurements for benchmark point 3. The fitted Gaussian distribution
  has the form $g(m_0, \, m_{1/2}) = p_0 \cdot \exp[-0.5/(1 - p_5^2)
  \cdot ([(m_0 - p_1)/p_2]^2 + [(m_{1/2} - p_3)/p_4]^2 - 2 p_5/(p_2 \,
  p_4) \cdot (m_0 - p_1) \cdot (m_{1/2} - p_3))]$, i.e.~$p_1$ and
  $p_2$ are the mean value and standard deviation of $m_0$, $p_3$ and
  $p_4$ are the mean value and standard deviation of $m_{1/2}$. The
  correlation factor $p_5 \equiv \rho$ is given by
  eq.(\ref{equ:CorrelationCoefficient}). The ANN for $m_0$ contains 15
  hidden neurons and underwent $244$ learning steps with a final control
  error of $0.106$, and the one for $m_{1/2}$ contains 20 hidden neurons
  and underwent $107$ learning steps with a final control error of
  $0.136$. The true values are $m_0 = \unit[1000]{GeV}$ and $m_{1/2} =
  \unit[600]{GeV}$.}
\label{fig:Gaussian2}
\end{figure*}

This is illustrated in Fig.~\ref{fig:Gaussian2}, which shows the
two--dimensional distribution of reconstructed $(m_0, \, m_{1/2})$
values for benchmark point 3. The left frame shows a
three--dimensional view and the right frame a two--dimensional contour
plot. The fit yields $m_0 = \unit[(1034 \pm 42.3)]{GeV}$ and $m_{1/2}
= \unit[(607.5 \pm 9.3)]{GeV}$; the result for $m_{1/2}$ agrees with
that of Fig.~\ref{fig:Gaussian1}, as expected. There is a very mild
negative correlation between these two parameters, with correlation
coefficient $\rho_{m_0 \, m_{1/2}} = -0.291$. This coefficient is
related to the entries of the covariance matrix through the formula
\begin{equation} \label{equ:CorrelationCoefficient}
\rho_{m_0 \, m_{1/2}} = \frac{{\rm cov}(m_0,\, m_{1/2})}{\sigma_{m_0}
  \, \sigma_{m_{1/2}}}. 
\end{equation}
Equivalently, the angle $\phi$ between the major axis of the error
ellipse and the $x-$axis in the right frame of
Fig.~\ref{fig:Gaussian2} is:
\begin{equation} \label{equ:EllipseAngle}
\tan(2 \phi) = \frac{2 \, \rho_{m_0 \, m_{1/2}} \, \sigma_{m_0} \,
  \sigma_{m_{1/2}}}{|\sigma_{m_0}^2 - \sigma_{m_{1/2}}^2|} 
\end{equation}
The ellipse is tilted with $\phi = \unit[-3.84]{^\circ}$ to the right bottom. 

The second method of determining errors on the parameter estimates of
the ANNs uses Gaussian error propagation. Since we know the covariance
matrix of the input values and the neural network functions we can
directly calculate the variances and covariances of the output
values. The variance of the output value $r \in \{m_0, \, m_{1/2}, \,
\tan\beta, \, A_0\}$ is then:
\begin{equation} \label{equ:error}
\sigma^2_r = \sum\limits_{i=1}^{N_O} \left( \frac{\partial
    f^*_r}{\partial O_i} \, \sigma(O_i) \right)^2 + 2 \cdot
\sum\limits_{i=1}^{N_O-1} \sum\limits_{j=i+1}^{N_O}
\frac{\partial f^*_r}{\partial O_i} \, \frac{\partial f^*_r}{\partial
  O_j} \, {\rm cov}(O_i, O_j) \,.
\end{equation}
Here $\sigma(O_i)$ is the standard deviation of the input observable
$O_i$ and ${\rm cov}(O_i, O_j)$ is the covariance between the
observables $O_i$ and $O_j$; explicit expressions for these quantities
are listed in Sec.~\ref{sec:Observables}. The sums run only over those
observables that have not been set to zero when we removed observables
with large statistical uncertainty as described in
Sec.~\ref{sec:Improvement}, i.e.~$N_O$ is in practice significantly
smaller than $N_{O,{\rm max}} = 84$. Finally, $f^*_r$ is the function
mapping the (unnormalized) observables onto the estimated CMSSM
values. Its derivative has the form:\footnote{Since each ANN has only
  one output, we removed the first index $r$ on the weights
  $w^{(2)}_{ra}$ in the second layer, cf. eq.(\ref{equ:outputNeuronInput}).}
\begin{eqnarray}    
\frac{\partial f^*_r}{\partial O_i} & = & \frac{\partial x_i}{\partial
  O_i} \, \frac{\partial f^*_r}{\partial f_r} \, \frac{\partial
  f_r}{\partial x_i} = \frac{\sigma^{tg}_r}{\sigma^{tg}_i} \cdot
\frac{\partial f_r}{\partial x_i} \nonumber\\ 
& = & \frac{\sigma^{tg}_r}{\sigma^{tg}_i} \cdot \left[
  \sum\limits^v_{a = 1} w_a^{(2)} \, w_{ai}^{(1)} \, \left( 1 -
    \tanh^2(\sum\limits^{N_O}_{j = 1} w_{aj}^{(1)} \, x_j +
    w_{a0}^{(1)}) \right) \right] \, .
\end{eqnarray}
Recall that $v$ is the number of hidden neurons. Moreover, as
described in Sec.~\ref{sec:Normalization} the ANNs, described by the
functions $f_r$, are trained to produce inversely normalized output
values from normalized inputs; the corresponding derivatives $\partial
x_i / \partial O_i$ and $\partial f^*_r / \partial f_r$ can be read
off eqs.(\ref{equ:InputNormalization}) and
(\ref{equ:OutputNormalization}), respectively. The training set
standard deviation $\sigma^{tg}_i$ for the $i$--th input value is
calculated using equation (\ref{equ:VarianceInputNormalization}) and
similarly $\sigma^{tg}_r$ for the output value $r$. The $x_i$ are the
normalized input values of the (simulated) measurement; again only the
$N_O$ inputs that have not been set to zero need to be taken into
account.

Similarly the covariance between two CMSSM values $r$ and $s$ is
\begin{equation} \label{equ:covariance}
{\rm cov}(r, \, s) = \sum\limits_{i=1}^{N_O} \frac{\partial
  f^*_r}{\partial O_i} \, \frac{\partial f^*_s}{\partial O_i} \,
\sigma^2(O_i) + \sum\limits_{i=1}^{N_O-1} \sum\limits_{j=i+1}^{N_O}
 {\rm cov}(O_i, \, O_j) \cdot \left( \frac{\partial
    f^*_r}{\partial O_i} \frac{\partial f^*_s}{\partial O_j} +
  \frac{\partial f^*_r}{\partial O_j} \frac{\partial f^*_s}{\partial
    O_i} \right). 
\end{equation}
It is reassuring to note that the covariances derived in this way are
close to the values obtained by fitting Gaussians to the distributions
of reconstructed CMSSM parameters. For example, for benchmark point 3
we find $\sigma(m_0) = \unit[47.3]{GeV},\, \sigma(m_{1/2}) =
\unit[11.5]{GeV}$ and $\rho_{m_0 \, m_{1/2}} = -0.309$; recall that
the corresponding values derived from the Gaussian fits are
$\unit[42.3]{GeV},\, \unit[9.3]{GeV}$ and $-0.291$, respectively. We
will see later that the two methods agree even more closely if a
higher integrated luminosity is used for the ``measurement''.

\subsubsection{Number of Hidden Neurons}
\label{sec:Hidden}
\setcounter{footnote}{0}

In order to determine the appropriate number of hidden neurons for
each ANN we start with a low number (like 10) and train the
corresponding network as described in Sec.~\ref{sec:Learning}. Note
that each additional hidden neuron adds $85 + 1 = 86$ free parameters
to the description of ANNs with one output neuron. As discussed in the
Appendix, the training of the ANN involves the repeated computation of
a matrix whose dimension grows linearly with $v$. For a fixed number
of learning steps the time needed to train an ANN therefore scales
like $v^2$. Hence the number of hidden neurons should not be increased
needlessly.

After the training, the distribution of estimated outputs for the
given benchmark point are calculated as described in the first part of
Sec.~\ref{sec:Error}.  If this distribution is (approximately) a
Gaussian the number of hidden layer neurons should be sufficient. On
the other hand, if the distribution does not look very Gaussian the
number of hidden layer neurons is increased and the process is
iterated, until a satisfactory result is achieved.\footnote{The ANNs
  were often not trained until the global minimum of the control error
  was reached before looking at the distribution of output
  values. Instead this distribution was checked already when the first
  local minimum of the control error was reached; if this distribution
  looked very non--Gaussian, the ANN was discarded and $v$ was
  increased.}

This completes our discussion of the construction of the ANNs. Before
reporting our results for the four benchmark points, we describe an
alternative strategy to determine the CMSSM parameters, based on
$\chi^2$ minimization.

\subsection{$ \chi^2$ Minimization}
\label{sec:ChiSquaredMinimization}
\setcounter{footnote}{0}

The preceding discussion shows that quite a lot of (computational)
effort is required to construct ANNs that are able to produce reliable
estimates of CMSSM parameters from simulated measurements.  Moreover,
a trained ANN is a ``black box''; it is very difficult to get a
feeling for how the input affects the output.

We therefore also attempted to derive estimates for the CMSSM
parameters by minimizing a $\chi^2$ function. At least conceptually
this is far simpler than training ANNs. However, it turns out that
fluctuations due to the finite Monte Carlo statistics used in deriving
the predictions of the model make it very difficult to derive
estimates for the parameters with statistically meaningful errors in
this way. In this subsection we describe the method we used for the
$\chi^2$ minimization and show the results for benchmark point 4.

We wish to minimize the $\chi^2$ between the simulated measurement and
predictions, which is given by
\begin{equation} \label{equ:chi2MP} \chi^2_{MP} =
  \sum\limits_{i,j=1}^{N_O^\prime} (O_i^M - O_i^P) V^{-1}_{ij} (O_j^M
  - O_j^P) \, .
\end{equation}
$O_i^M$ are the observables of the simulated measurement and $O_i^P$
are the predictions for these observables derived for a particular set
of CMSSM parameters. The double sum runs over all observables that
have been derived from a minimum number of events (see below), and $
V^{-1}_{ij}$ is the inverse covariance matrix of the relevant
observables, with entries
\begin{equation} \label{equ:Vmn}
V_{ij} = {\rm cov}(O^M_i, O^M_j) + {\rm cov}(O^P_i, O^P_j) \,.
\end{equation}
The second term in eq.(\ref{equ:Vmn}) takes into account the
statistical error on the prediction due to the finite size of the
Monte Carlo sample that has been used for its calculation. In
\cite{bod4} it has been shown that this function has the statistical
properties of a true $\chi^2$ distribution, provided we only include
observables that have been computed from sufficiently many events.  In
particular, for $\unit[10]{fb^{-1}}$ of simulated integrated
luminosity the total number of events is taken into account if the
measurement or the prediction has at least one event after cuts; for
our benchmark points this is always the case. The event ratio $n_c/N =
O_{1,c}$ for a class $c$ is included in the definition of $\chi^2$ if
the class contains at least $10$ events for the measurement {\em or}
the prediction. All other observables $O_{i,c}, \ i \geq 2$ are only
included if $n_c \geq 10$ for both the measurement {\em and} the
prediction.\footnote{Some of these minimal event numbers differ from
  the ones used in \cite{bod4}. The benchmark points investigated here
  yield on the order of $O(1000)$ events after cuts for
  $\unit[10]{fb^{-1}}$ of data, as shown in Sec.~2.2. In contrast the
  parameter sets analyzed in \cite{bod4} yielded on average around
  $25,000$ events after cuts. Using $10$ as minimal $n_c$ for all
  classes showed a little bit better results than the old choice of
  $10$ for $n_c/N$ and $\langle H_T \rangle_c$, $50$ for $n_{c,b}/n_c$
  and $\langle j \rangle_c$, and $500$ for $n_{c,\tau^+}/n_c$ and
  $n_{c,\tau^-}/n_c$.} For an higher integrated luminosity the thresholds
are scaled up accordingly.

Depending on the number of events that are generated, i.e.~the assumed
integrated luminosity, computing the prediction for a single set of
CMSSM parameters can already be quite costly in terms of CPU time. We
therefore perform the $\chi^2$ minimization in two steps. The first
step is supposed to roughly identify the correct region of parameter
space in which the minimum lies; the second step should then pin down
the location of the minimum. In both steps we compute the predictions
using $\unit[10]{fb^{-1}}$ of simulated data.

In the first step we use an algorithm based on simulated annealing.
We begin by computing the prediction for a randomly chosen set of
parameters, and compute the corresponding $\chi^2$. We then randomly
vary one of the CMSSM parameters, leaving the other three parameters
unchanged. If the resulting $\chi^2$ is smaller than the previous one,
the corresponding parameter set is taken as the new best guess for the
location of the minimum. If the new $\chi^2$--value is bigger than the
previous one, there is still a finite probability that the new
parameters are picked as the new best guess; this probability
decreases exponentially with the difference between the two $\chi^2$
values. This approach should prevent the algorithm from getting stuck
in a local minimum. The same parameter would then be changed again, if
a new minimum was not picked; this loop is terminated if a given
number of tries did not improve the minimization. After that the
second parameter is changed and so on. When all four parameters have
been scanned in this manner, the scanning starts at the first
parameter again. The procedure is continued until a given number of
steps (here 250) is reached.

Recall that our goal in the first step is just to get a rough estimate
of the CMSSM parameters. Therefore, we did not put much effort into setting the
available parameters of the simulated annealing algorithm.
Nevertheless this step is useful, in particular for narrowing down the
range for $m_0$. The size of $m_0$ approximately determines the
maximal allowed range of $A_0$. This is important for the second step
of the minimization, in which the parameters should be determined more
exactly.

In this second step we use the $\chi^2$ minimization algorithms
``Simplex'' and ``Migrad'' of TMinuit \cite{minuit} in the program
ROOT. Using this algorithm we set allowed parameter ranges in order to
avoid parameter selections where simulation problems occur (e.g.~very
large $|A_0|$ can lead to problems with the generation of the
superparticle and Higgs spectrum). These algorithms will also work
better if the starting point is closer to the true minimum and its
distance to the minimum is roughly known. Therefore we use the output
of the simulated annealing algorithm as starting point; the $p$--value
corresponding to this starting point gives us a rough idea how close
we are to the true minimum.

These algorithms eventually do converge onto a new best guess for the
location of the minimum. Since Minuit has been designed specifically
for $\chi^2$ minimizations, it even gives estimates for the
statistical errors of the determined parameters. However, it turns out
that these error estimates are almost always much too small. This can
be traced back to the statistical fluctuations of the predictions. The
final minimum found is nearly always produced by a rather extreme
Monte Carlo fluctuation in the prediction. Such a fluctuation is
likely to occur only in a region of parameter space around the true
CMSSM parameters, i.e.~the location of the minimum is likely not too
far from the true location of the minimum computed from predictions
with negligible error. However, in our simulation the fluctuation
frequently reduced $\chi^2$ by several units, leading to a very steep,
but spurious, minimum. The width of this spurious minimum, which is
what Minuit attempts to estimate, is then a very poor estimate for the
actual error on the CMSSM parameters.

This problem can be ameliorated by increasing the integrated
luminosity used for the computation of the predictions, which reduces
the fluctuations. Alas, these fluctuations only decrease inversely to
the square root of the number of generated events, whereas the CPU
time needed to generate them obviously scales linearly with this
number. Note that Minuit needs several hundred search steps to
converge on a minimum, i.e.~several hundred sets of CMSSM parameters
have to be simulated for each $\chi^2$ minimization by Minuit. The
available time then limits the number of events we can generate for
each parameter set. We found that even using an integrated luminosity
of $\unit[500]{fb^{-1}}$ instead of $\unit[10]{fb^{-1}}$ for the
predictions did not solve the problem of the extreme statistical
fluctuations leading to too small errors.

We therefore derive final estimates for the CMSSM parameters, and
their errors, by simulating up to 500 parameter sets (on several CPUs)
with an integrated luminosity of $\unit[500]{fb^{-1}}$ around the
minimum found by TMinuit, and calculate the corresponding
$\chi^2$--values.\footnote{Of course, the predicted number of events
  is divided by $50$ when comparing a $\unit[500]{fb^{-1}}$ prediction
  to a $\unit[10]{fb^{-1}}$ measurement.} These points are then used
to fit $\chi^2$ as a quadratic function of $m_0, \, m_{1/2}, \,
\tan\beta$ and $A_0$:
\begin{eqnarray} \label{equ:chifit}
\chi^2 & = & \chi^2_{\rm min} + \Delta\chi^2 \nonumber \\
 & = & \chi^2_{\rm min} + \begin{pmatrix} m_0 - m_0^{\rm min}\\m_{1/2} -
   m_{1/2}^{\rm min}\\ \tan\beta - \tan\beta^{\rm min} \\ A_0 -
   A_0^{\rm min} \end{pmatrix}^T \cdot V^{-1} \cdot \begin{pmatrix} m_0 -
   m_0^{\rm min} \\ m_{1/2} - m_{1/2}^{\rm min} \\ \tan\beta -
   \tan\beta^{\rm min} \\ A_0 - A_0^{\rm min} \end{pmatrix} \,.
\end{eqnarray}
Here $m_0^{\rm min}, \, m_{1/2}^{\rm min}, \,\tan\beta^{\rm min}, \,
A_0^{\rm min}$ are our final estimates of the CMSSM parameters from
the $\chi^2$ minimization, $\chi^2_{\rm min}$ is the corresponding
minimal $\chi^2$--value and $V^{-1}$ is the final estimate for the
inverse covariance matrix of the extracted parameters, from which we
compute the errors on the estimated parameters including their
correlations. Note that eq.(\ref{equ:chifit}) contains $15$ free
parameters ($\chi^2_{\rm min}$, the values of the CMSSM parameters,
and the entries of the covariance matrix, which is a symmetric $4
\times 4$ matrix); we again use ``Simplex'' and ``Migrad'' in TMinuit
to determine them. The large number of parameter sets used in this
fit, as well as the large integrated luminosity used for each
parameter set, should reduce the effect of statistical fluctuations in
the prediction. Specifically, we determine the free parameters of
eq.(\ref{equ:chifit}) by minimizing the summed differences
\begin{equation} \label{equ:chiSquaredFit}
\sum\limits_{n = 1}^{500} \frac{\left[ \chi^2_n - \chi^2(m_0^{(n)}, \,
  m_{1/2}^{(n)}, \, \tan\beta^{(n)}, \, A_0^{(n)}) \right]^2}
{(\chi^2_n)^d} \, .
\end{equation}
Here $\chi^2_n$ is the $\chi^2$--value of parameter set $n$ and
$\chi^2(m_0^{(n)}, \, m_{1/2}^{(n)}, \, \tan\beta^{(n)}, \,
A_0^{(n)})$ is the calculated $\chi^2$ which has been computed as in
eq.(\ref{equ:chifit}) from the current values of the fit
parameters. Finally, the parameter $d$ determines the weight of
parameter sets that are far away from the $\chi^2$ minimum; $d = 0$
means that all $500$ parameter sets have equal weight, whereas a
positive $d$ suppresses the weight of parameter sets with high
$\chi^2$--values which should be farther away from the minimum. Note
that $\chi^2$ can be expected to be a quadratic function of the CMSSM
parameters only in the vicinity of its minimum; further away $\chi^2$
may change quite quickly, e.g.~if some new decay modes open up. Using
the functional form (\ref{equ:chifit}) also for parameter sets which
are far away from the minimum may therefore distort the fit. This
argues in favor of using a positive $d$.

Unfortunately we find that the choice of $d$ has a relatively big
influence on the outcome of the fit. We therefore need a criterion to
determine the optimal choice of $d$. Recall that in the ANN analysis
we determined the number of hidden neurons by requiring that a
Gaussian distribution of input sets also produces a Gaussian
distribution of output values. We want to apply the same criterion
here. Note, however, that for each set of measurements we need to
re--do the fit of eq.(\ref{equ:chifit}); since $15$ parameters need to
be determined in each fit, it takes some amount of CPU
time.\footnote{Strictly speaking we would have to re--do the entire
  $\chi^2$ minimization for each version of the simulated measurement,
  including the first two steps. However, the region of parameter
  space populated by the $500$ sets of predictions used in
  eq.(\ref{equ:chifit}) is sufficiently large that it should include
  the results of the first two steps as applied to the variants of the
  original measurement. Since steps 1 and 2 require additional event
  generation, and are thus very time consuming, re--doing the entire
  $\chi^2$ minimization for these sets was in any case not a realistic
  option for us.} We therefore ``only'' create $1,000$ Gaussian
distributed variants of the measurement (rather than $100,000$ in the
ANN analysis) and compare each of these ``data sets'' with all
predictions. This leads to $1,000$ sets of fitted CMSSM parameters
$m_0^{\rm min}$, $m_{1/2}^{\rm min}$, $\tan\beta^{\rm min}$ and
$A_0^{\rm min}$. The distributions of those CMSSM parameters should
then also be Gaussians. We therefore use the form of these
distributions to check the reliability of our $\chi^2$ minimization
method.

\begin{figure*}[h!]
\centering
\resizebox{0.94\textwidth}{!}{
\includegraphics{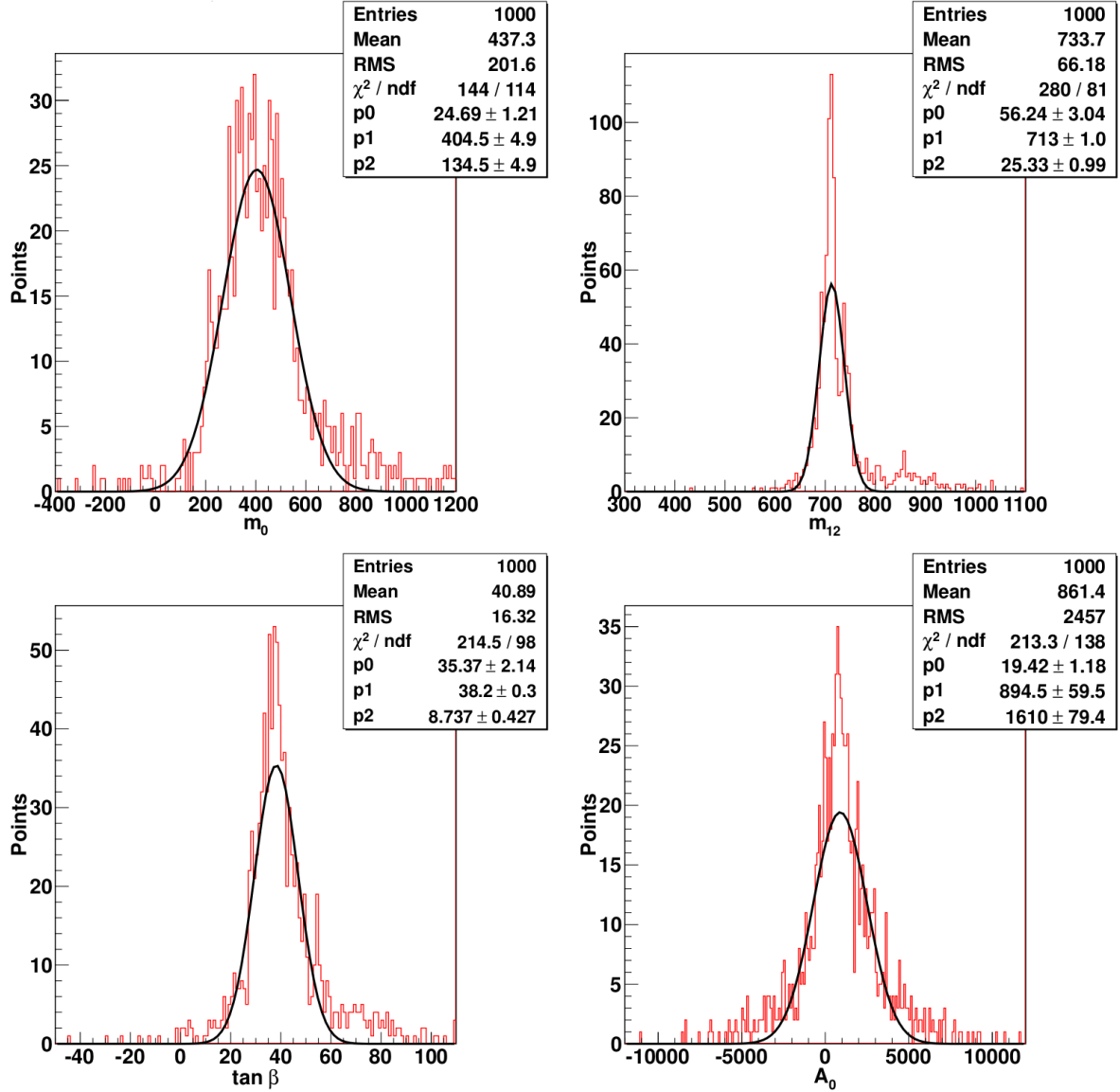}
}
\caption{Distributions of output values of the $\chi^2$ minimization
  for benchmark point 4, for a simulated measurement with integrated
  luminosity of $\unit[10]{fb^{-1}}$. $m_0$, $m_{1/2}$ and $A_0$ are
  given in GeV. $1,000$ Gaussian distributed versions of the original
  simulated measurement are fed into (\ref{equ:chiSquaredFit}) with $d
  = 3$ to determine $1,000$ sets of CMSSM parameters. The
  distributions of these parameters (histograms in the four frames)
  are fitted to Gaussians (curves). ``Mean'' and ``RMS'' are the
  usual mean value and standard deviation of the entries in the
  histograms. The other parameters are derived from the Gaussian fits,
  $g(x) = p_0 \cdot \exp(-1/2 [(x - p_1)/p_2]^2)$ with $x$ being the
  parameter in question. The $x-$axes are centered on the input values
  of these parameters, $m_0 = \unit[400]{GeV}$, $m_{1/2} =
  \unit[700]{GeV}$, $\tan\beta = 30$ and $A_0 = 0$.}
\label{fig:Fehler4}
\end{figure*}

An example of this check is shown in Fig.~\ref{fig:Fehler4}, which
depicts the situation for benchmark point 4 and a simulated
measurement with integrated luminosity of $\unit[10]{fb^{-1}}$. Here
$d=3$ has been used in eq.(\ref{equ:chiSquaredFit}). We see that the
distributions of all four CMSSM parameters have significant
non--Gaussian tails. These become even more prominent if we only use
$300$, rather than $500$, predictions for the final fit of the CMSSM
parameters in eq.(\ref{equ:chiSquaredFit}); this indicates that the
distribution of reconstructed CMSSM parameters might become somewhat
more Gaussian if more than $500$ predictions are used. However, the
change between the results based on $300$ and $500$ predictions is not
dramatic, so it seems unlikely that properly Gaussian distributions of
the extracted CMSSM parameters can be obtained from a number of
predictions that can be generated (by us) within a reasonable amount
of time. We conclude that, while the central values of the derived
CMSSM parameters are within one estimated standard deviation from
their true values, the error estimates are not reliable.

\begin{figure*}[h!]
\centering
\resizebox{0.7\textwidth}{!}{
  \includegraphics{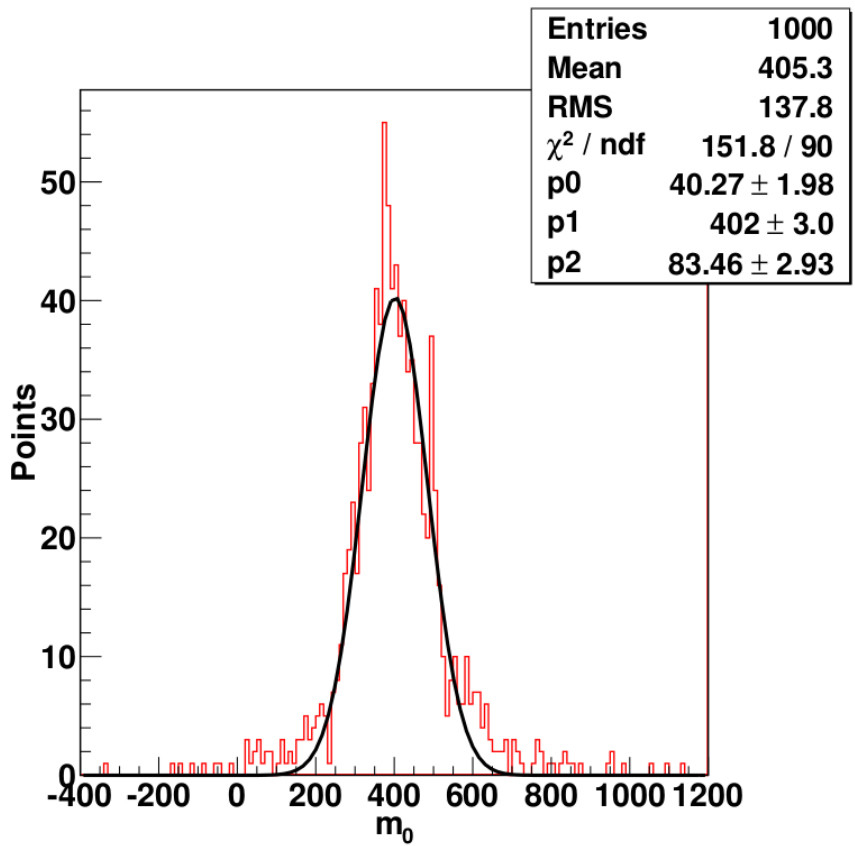}
}
\caption{The same distribution as in Fig.~\ref{fig:Fehler4}, but based
  on a simulated measurement with $\unit[500]{fb^{-1}}$ of data.}
\label{fig:Fehler50}
\end{figure*}

This is also indicated by Fig.~\ref{fig:Fehler50}, which shows the
distribution of reconstructed $m_0$ values for a simulated measurement
with fifty times higher integrated luminosity. Under the usual scaling
law, the statistical error should then decrease by a factor $\sqrt{50}
\simeq 7$. Instead our procedure gives an error which is less than a
factor of two smaller than that for $\unit[10]{fb^{-1}}$ of data. Note
that we have increased the minimal number of events in a given class
we require if the corresponding observables are to be included into
the overall $\chi^2$ also by a factor of fifty, i.e.~observable
$O_{i,c}$ is included only if $n_c \geq 500$. This should ensure that
Gaussian statistics should be applicable everywhere. In
Fig.~\ref{fig:Fehler50} we nevertheless still see substantial
non--Gaussian tails.

\begin{figure*}[h!]
\centering
\resizebox{1.\textwidth}{!}{
 \includegraphics{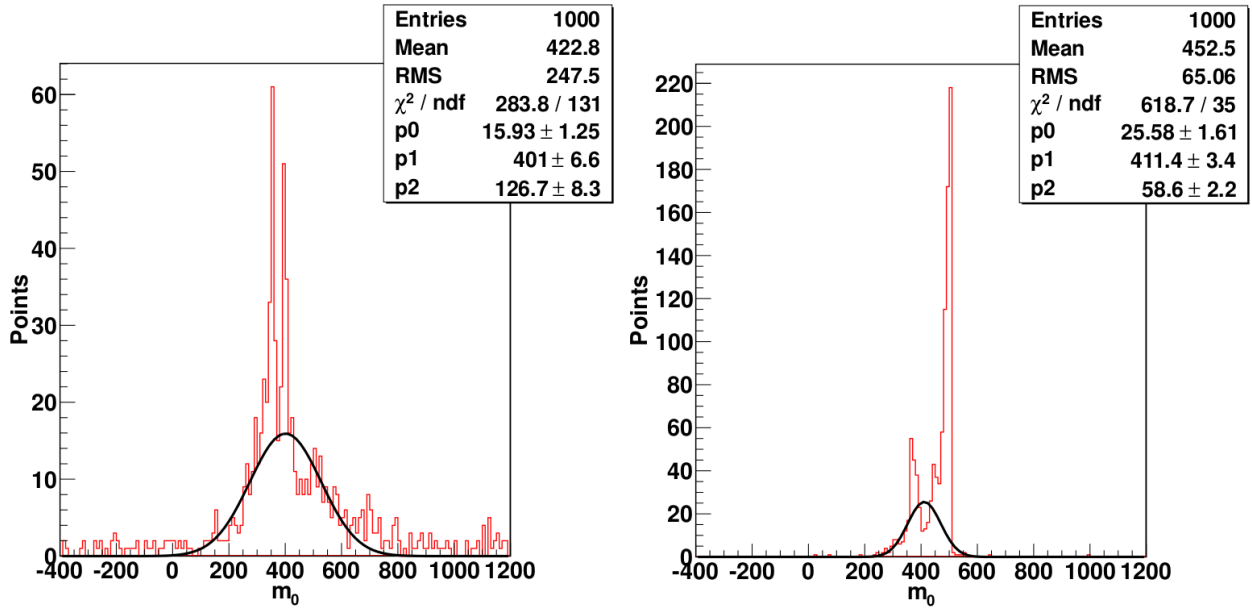}
}
\caption{The same distribution as in Fig.~\ref{fig:Fehler50}, but using
  $d = 2$ (left) and $d = 4$ (right) in eq.(\ref{equ:chiSquaredFit}),
  instead of $d = 3$.}
\label{fig:Fehler2_4}
\end{figure*}

Another problem is that the results of this method depend sensitively
on the power $d$ in eq.(\ref{equ:chiSquaredFit}). This can be seen
from Fig.~\ref{fig:Fehler2_4}, which shows the distribution of
reconstructed $m_0$ values for the same simulated measurements based on
$\unit[500]{fb^{-1}}$ used in Fig.~\ref{fig:Fehler50}, but using $d =
2$ and $d =4$ instead of $d = 3$. Evidently these distributions do no
longer have much resemblance to Gaussians. Moreover, the estimated
error on $m_0$ differs by a factor of two between these two choices.

As noted above, the fit based on eqs.(\ref{equ:chifit}) and
(\ref{equ:chiSquaredFit}) which yields the estimates of the CMSSM
parameters also provides estimates of the statistical errors on these
extracted parameters. While the errors estimated in this way
generally are of the same order of magnitude as the errors obtained in
the Gaussian fits shown in Figs.~\ref{fig:Fehler4} to
\ref{fig:Fehler2_4}, the numerical values of the errors fluctuate
quite widely within the sample of $1,000$ versions of the
``measurement'' shown in these figures. In fact, the estimates of the
errors fluctuate much more than the central values do.

We finally note that the error estimates derived from the $\chi^2$
minimization, while quite uncertain, are generally significantly
larger than those found by the ANNs, which {\em are} reliable as
illustrated above. The $\chi^2$ fit typically gives two to three times
bigger errors when using simulated measurements based on
$\unit[10]{fb^{-1}}$ of data; for simulated measurements assuming an
integrated luminosity of $\unit[500]{fb^{-1}}$ the error estimates
from the $\chi^2$ fits are as much as a full order of magnitude larger
than those of the ANNs. 

We therefore conclude that a $\chi^2$ fit yielding reliable results for
both the central values and the statistical errors of the CMSSM
parameters would need significantly {\em more} computational effort
than the ANNs, in order to reduce the Monte Carlo errors on the
predictions to a level that can be tolerated by this method. Moreover,
while the error estimates we presented above are not yet reliable,
they do suggest that the precision of the parameter estimates from
$\chi^2$ fitting will be significantly worse than that of the ANNs. We
did therefore not attempt to further optimize the $\chi^2$ fits, and
instead present the results for the ANNs for all four benchmark points.

\section{Results}
\label{sec:Results}
\setcounter{footnote}{0}

Recall that we constructed a separate ANN, with a single output
neuron, for each CMSSM parameter. Moreover, as described in
Sec.~\ref{sec:Sets}, we generated different training and control sets
for each benchmark point. Altogether we thus ended up constructing $4
\times 4 = 16$ different ANNs. 

Some details of these $16$ ANNs are collected in
Table~\ref{tab:performance}. We need between $15$ and $25$ hidden
neurons to construct ANNs that produce an approximately Gaussian
distribution of output values when fed with a Gaussian distribution of
inputs. Here we already note a first different between the ANNs
designed for different CMSSM parameters: the ones for $m_0$ and
$m_{1/2}$ usually need slightly fewer hidden neurons than the ones for
$A_0$ and $\tan\beta$.

\begin{table*}[h!]
\centering
\caption{Overview of the $16$ ANNs we constructed: number of hidden
  neurons, number of learning steps and final control errors. These
  results are based on simulated measurements for an integrated
  luminosity of $\unit[10]{fb^{-1}}$. The control error is normalized
  as in Eq.(\ref{equ:NormalizedError}).} 
\vspace*{3mm}
\begin{tabular}{|cl||c|c|c|}
\hline
 & & No. of hidden neurons & No. of learning steps & Control error\\
\hline
 & $m_0$ & $20$	& $1009$	& $0.101$ \\ 
Reference & $m_{1/2}$& $20$ & $407$ & $0.081$ \\ 
Point 1	& $\tan\beta$ & $20$ & $653$ & $0.135$ \\
 & $A_0$ & $20$	& $561$	& $0.532$ \\
\hline
 & $m_0$ & $20$	& $929$	& $0.123$ \\ 
Reference & $m_{1/2}$ & $15$ & $452$ & $0.116$ \\ 
Point 2	& $\tan\beta$ & $20$ & $499$ & $0.387$ \\
 & $A_0$ & $25$	& $155$ & $0.711$ \\
\hline
 & $m_0$ & $15$	& $244$ & $0.106$ \\ 
Reference & $m_{1/2}$ & $20$ & $107$ & $0.136$ \\ 
Point 3	& $\tan\beta$ & $20$ & $240$ & $0.681$ \\
 & $A_0$ & $15$ & $763$ & $0.405$\\
\hline
 & $m_0$ & $22$ & $488$ & $0.156$ \\ 
Reference & $m_{1/2}$ & $22$ & $637$ & $0.067$ \\ 
Point 4 & $\tan\beta$ & $25$ & $511$ & $0.178$ \\
 & $A_0$ & $25$	& $391$	& $0.310$ \\
\hline
\end{tabular}
\label{tab:performance}
\end{table*}

Table~\ref{tab:performance} also shows that the control error reaches
its minimum after typically hundreds of learning steps. We continued
the training usually for about $100$ learning steps after each
minimum, in order to check that the absolute minimum has indeed been
found. Moreover, as shown in the Appendix, each learning step of an
ANN with $v$ hidden neurons entails the calculation of a matrix with
$86 \cdot v + 1$ rows and columns; the computation of each element of
this matrix requires a sum over all $\sim 100,000$ members of the
training set for this ANN. As a result, the training of a single ANN
can take up to $2,000$ CPU hours.\footnote{In particular for the ANNs
  with high numbers of completed learning steps, the last hundreds of
  learning steps led only to relatively small improvements of the
  control errors and calculated standard deviations. Therefore, for
  those ANNs nearly as good results can be reached with much less
  computational effort.}

The last column of Table~\ref{tab:performance} gives the normalized
control error, as defined in Eq.(\ref{equ:NormalizedError}). The size
of this error should determine the accuracy with which a given CMSSM
parameter can be determined by its ANN. We see that this error is
significantly smaller for $m_0$ and $m_{1/2}$ than for $A_0$ and
$\tan\beta$. This is not surprising, since the former parameters affect
the spectrum of superparticles, in particular the spectrum of strongly
interacting superparticles, more than the latter do. We also see that
the normalized control error is usually somewhat smaller for the
gaugino mass $m_{1/2}$ than for scalar mass $m_0$; the one exception
is benchmark point 3, where $m_0$ is large enough to significantly
affect the squark masses (in contrast to benchmark point 1, where
$m_0$ essentially only affects the slepton masses), but not so large
that the cross section for the production of final states involving
one or two squarks is much smaller than that for gluino pair
production (in contrast to benchmark point 2). Note also that the
normalized control error for $m_{1/2}$ is smaller for points 1 and 4,
where charginos and neutralinos have competing two--body decays into
sleptons and into gauge or Higgs bosons, than for points 2 and 3,
where charginos and neutralinos cannot decay into sleptons. Sizable
branching ratios into sleptons increase the fraction of events
containing several charged electrons or muons, and/or increase the
average number of reconstructed $\tau$'s in the final state. These
branching ratios therefore affect several of our observables.

The normalized control error for $\tan\beta$ differs quite
significantly between the four benchmark points. It is smallest for
point 1, where the kinematics and branching fractions for two--body
decays of gluinos into $\tilde b$ squarks, and of neutralinos and
charginos into $\tilde \tau$ sleptons, depend on $\tan\beta$. In
contrast, in point 2 gluinos only have three--body decays, and the
neutralinos and charginos cannot decay into sleptons, leading to a
greatly reduced sensitivity to $\tan\beta$. In point 3, all gluinos
decay into $\tilde t_1$, whose mass and branching ratios depend only
weakly on $\tan\beta$ unless the latter is very large. Also,
neutralinos and charginos again cannot decay into sleptons here. As a
result, this point has the largest normalized control error on
$\tan\beta$. Finally, point 4 has a rather large input value of
$\tan\beta = 30$. As a result, terms proportional to $m_b \tan\beta$
and $m_\tau \tan\beta$ in the RG equations as well as sfermion mass
matrices are quite sizable, yielding significant effects in some of
our observables. The normalized control error on $\tan\beta$ is
therefore again relatively small for this point.

The ANNs for $A_0$ typically have the largest normalized control
errors. At the tree level, $A_0$ practically only affects the $\tilde
t$ sector; $L-R$ mixing in the $\tilde b$ and $\tilde \tau$ sectors is
dominated by a term $\propto \mu \tan \beta$, which is much larger than
the relevant (weak--scale) $A$ parameter for all our benchmark
points. $A_0$ also affects third generation sfermion and Higgs boson
masses via the RG running, and hence also the value of $|\mu|$ which
is determined from the requirement that the mass of the $Z$ boson
comes out correctly. From dimensional analysis and the structure of
the RG equations, any weak--scale soft breaking mass can be written as
\begin{equation} \label{equ:mi}
m_i^2 = a_i m_0^2 + b_i m_{1/2}^2 + c_i A_0^2 + d_i m_{1/2} A_0\,;
\end{equation}
the coefficients depend on the dimensionless couplings in the theory,
and hence also on $\tan\beta$. A very similar equation therefore also
holds for the derived weak--scale value of $\mu^2$.\footnote{There is
  also a contribution $\propto M_Z^2$, which is however much smaller
  than term $\propto m_{1/2}^2$ in the CMSSM.} Benchmark points 1, 2
and 4 have $A_0 = 0$, so the sensitivity to $A_0$ via the quadratic
coefficients $c_i$ is small. Moreover, the $d_i$ are typically
significantly smaller in magnitude than $\max\{|a_i|,\, |b_i|\}$,
further reducing the sensitivity to $A_0$. On the other hand,
benchmark point 3 does have a large $A_0$; moreover, we saw that this
point has an especially large normalized control error for
$\tan\beta$. As a result, benchmark point 3 actually has a smaller
normalized control error for $A_0$ than for $\tan\beta$.  Altogether,
the biggest error for $A_0$ ($0.711$, for benchmark point 2) is over
ten times larger than the smallest error for $m_{1/2}$ ($0.067$, for
benchmark point 4).

In fact, the entries of Table~\ref{tab:performance} underestimate the
differences between the relative size of the errors on the CMSSM
parameters. The reason is that the control error is normalized, as
shown in Eq.(\ref{equ:NormalizedError}); it thus scales inversely with
the size of the region covered by the respective training set. We saw
in Sec.~\ref{sec:Sets} that these sets only cover a range of
$\unit[80]{GeV}$ in $m_{1/2}$ and $250$ to $\unit[350]{GeV}$ in $m_0$,
but much of the theoretically allowed parameter space for $\tan\beta$
and $A_0$. We thus expect the final errors on the former two
parameters to be much smaller than those on the latter two.

\begin{table}[h!]
\centering
\caption{Values of the CMSSM parameters reconstructed by our ANNs for
  the four benchmark points, their uncertainties, and the corresponding
  correlation coefficients. The standard deviations and correlation
  coefficients are calculated via error propagation, see
  eqs.(\ref{equ:error}) to (\ref{equ:covariance}). The dimensionful
  parameters $m_0$, $m_{1/2}$ and $A_0$ are given in GeV.}
\vspace*{3mm}
\begin{tabular}{|c||c|c|c|c|}
\hline
 & Point 1 & Point 2 & Point 3 & Point 4 \\
\hline 
 & \multicolumn{4}{c|}{10/fb} \\
\hline
 $\overline{m}_0 \pm \sigma_{m_0}$ & $ 167.25 \pm 33.78$ & $ 1998.93 \pm 92.20$ &
   $1055.97 \pm 47.26$ & $ 482.94 \pm 61.23$ \\
 $\overline{m}_{1/2} \pm \sigma_{m_{1/2}}$ & $ 697.51 \pm 7.36$ &
 $446.55 \pm 11.30$ & 
   $607.47 \pm 11.53$ & $695.48 \pm 7.87$ \\
 $\overline{\tan\beta} \pm \sigma_{\tan\beta}$ & $ 21.35 \pm 5.96$ & $
 9.67 \pm 18.81$ & 
   $23.41 \pm 37.42$ & $ 23.37 \pm 11.08$ \\
 $\overline{A}_0 \pm \sigma_{A_0}$ & $ 463.43 \pm 326.00$ & $1406.37
 \pm 2898.67$ & 
   $1453.49 \pm 1891.58$ & $-73.52 \pm 628.16$ \\
\hline
 $\rho_{m_0 \, m_{1/2}}$ & $-0.037$ & $-0.051$ & $-0.309$ & $-0.407$ \\
 $\rho_{m_0 \, \tan\beta}$ & $0.131$ & $-0.068$ & $0.044$ & $0.057$ \\
 $\rho_{m_0 \, A_0}$ & $0.021$ & $-0.251$ & $0.200$ & $-0.671$ \\
 $\rho_{m_{1/2} \, \tan\beta}$ & $0.201$ & $0.174$ & $-0.084$ & $-0.350$ \\
 $\rho_{m_{1/2} \, A_0}$ & $0.121$ & $-0.119$ & $0.065$ & $0.139$ \\
 $\rho_{\tan\beta \, A_0}$ & $-0.035$ & $0.249$ & $-0.218$ & $0.238$ \\
\hline
& \multicolumn{4}{c|}{500/fb} \\
\hline
 $\overline{m}_0 \pm \sigma_{m_0}$ & $ 156.23 \pm 4.86$ & $2004.05 \pm 10.61$ &
 $ 1015.66 \pm 4.49$ & $391.86 \pm 7.70$ \\
 $\overline{m}_{1/2} \pm \sigma_{m_{1/2}}$ & $ 701.05 \pm 0.87$ &
 $451.15 \pm 1.00$ & 
   $598.75 \pm 1.21$ & $700.71 \pm 0.88$ \\
 $\overline{\tan\beta} \pm \sigma_{\tan\beta}$ & $ 9.39 \pm 0.70$ &
 $14.66 \pm 2.17$ & 
   $20.79 \pm 5.20$ & $30.50 \pm 1.24$ \\
 $\overline{A}_0 \pm \sigma_{A_0}$ & $ -43.61 \pm 55.37$ & $261.47 \pm 474.68$ &
   $774.39 \pm 295.63$ & $183.05 \pm 101.48$ \\
\hline
 $\rho_{m_0 \, m_{1/2}}$ & $-0.125$ & $-0.519$ & $-0.550$ & $-0.695$ \\
 $\rho_{m_0 \, \tan\beta}$ & $-0.233$ & $-0.131$ & $0.294$ & $0.441$ \\
 $\rho_{m_0 \, A_0}$ & $0.038$ & $-0.068$ & $-0.557$ & $-0.232$ \\
 $\rho_{m_{1/2} \, \tan\beta}$ & $0.259$ & $0.076$ & $-0.110$ & $-0.416$ \\
 $\rho_{m_{1/2} \, A_0}$ & $0.033$ & $0.165$ & $0.509$ & $0.250$ \\
 $\rho_{\tan\beta \, A_0}$ & $-0.211$ & $-0.158$ & $-0.055$ & $0.464$ \\
\hline
\end{tabular}
\label{tab:errors}
\end{table}

This is confirmed by Table~\ref{tab:errors}, where we list the values
of the CMSSM parameters reconstructed from our ANNs fed with the
$\unit[10]{fb^{-1}}$ and $\unit[500]{fb^{-1}}$ simulated measurements
at the four benchmark points. The standard deviations and correlation
coefficients are calculated with the propagation of uncertainty method
as described at the end of Sec.~\ref{sec:Error}. If we divide the
final estimate of the standard deviation by the product of the
normalized control error of Table~\ref{tab:performance} and the size
of the parameter region spanned by the training sets, we obtain values
that cluster around $1.2 \pm 0.4$ ($0.15 \pm 0.04$) for the simulated
measurements with $\unit[10]{fb^{-1}}$ ($\unit[500]{fb^{-1}}$) of
data. In other words, the final estimate of the statistical
uncertainty is approximately proportional to this product. The main
exceptions to this rule occur for benchmark point 2, where the
estimated uncertainties for $m_0$ ($A_0$) are nearly two times bigger
than (less than half as big as) the value obtained from this simple
scaling. A strict scaling is not expected, since the normalized
control error measures the {\em average} performance of the ANN
against 300 CMSSM scenarios in the control set, whereas we are now
considering specific benchmark points. 

Moreover, the control error determines the deviation from the true
value, not the estimated size of the uncertainty of the extracted
CMSSM parameters. However, Table~\ref{tab:errors} also shows that the
estimated standard deviations reflect the differences to the true
values quite well. This is true in particular for the simulated
measurement based on $\unit[10]{fb^{-1}}$ of data, where $12$ of the
$16$ estimated parameter values are less than one estimated standard
deviation away from the true value, and the remaining $4$ estimates
differ from the true values by less than two estimated standard
deviations. 

Another indication for the reliability of the estimated standard
deviations is that they decrease approximately by the expected factor
$\sqrt{50} \simeq 7$ when going to the simulated measurement based on
$\unit[500]{fb^{-1}}$ of data. Recall that we only include statistical
uncertainties, which should of course decrease proportional to the
inverse square root of the accumulated luminosity.

However, upon closer expectation some systematic deviation from the
expected reduction of the estimated uncertainties by a factor of
$\sqrt{50}$ become apparent. First, when averaging over all four
benchmark points the estimated errors on $m_{1/2}, \, m_0$ and
$\tan\beta$ actually decrease by factors of $9.6, \, 8.5$ and
$8.3$. This indicates that the uncertainties for the
$\unit[500]{fb^{-1}}$ ``measurements'' might be somewhat
underestimated. In fact, Table~\ref{tab:errors} shows that in this
case $10$ out of $16$ reconstructed parameter values are more than one
estimated standard deviation away from their true values; $4$ of the
reconstructed values are more than two standard deviations off, and
one ANN output ($m_0$ for benchmark point 3) differs from its true
value by more than three estimated standard deviations. The likely
explanation for this is that the predictions in the training and
control sets were also ``only'' based on $\unit[500]{fb^{-1}}$ of
simulated data, i.e.~they had the same uncertainty due to finite Monte
Carlo statistics as our simulated measurements. This ``theoretical
uncertainty'' is not included in our estimates of the uncertainty of
the ANN outputs. Statistical fluctuations in the training and/or
control sets might also lead to systematic off--sets of the ANN
outputs relative to the true values. This might explain the relatively
poor performance of the ANNs for benchmark point 3 when fed the
simulated $\unit[500]{fb^{-1}}$ measurement, where all four output
values are more than one estimated standard deviation off, and three
of the four values are more than two estimated standard deviations
off. In order to check this, we simulated the $\unit[500]{fb^{-1}}$
measurement of benchmark point 3 two more times with different seeds
in Herwig++. For both additional versions the estimated value of $m_0$
was less than one standard deviation away from the true
value. Therefore, the more than three standard deviations in Table
\ref{tab:errors} seems to be a rather extreme statistical
fluctuation. On the other hand, for all three version of the
measurement, $A_0$ differed between two and three standard deviations
from the true value. The estimated value was always smaller than the
true value $A_0 = \unit[1500]{GeV}$. This slight tendency to lower
values might originate from the fact that the true value lies near the
upper bound $A_{0,{\rm max}}= 2 m_0$ enforced in the selection of the
training sets.

In contrast, the error estimates for the four values of $A_0$ only
decrease by a factor of $\sim 6.2$. For the reasons listed above, the
errors for the $\unit[500]{fb^{-1}}$ ``measurements'' might still be
slightly under--estimated. However, in the case at hand the errors for
the $\unit[10]{fb^{-1}}$ ``measurements'' are also somewhat suspect.
The reason is that in three cases the estimated ``$1 \, \sigma$''
interval for $A_0$ extends beyond the range covered in the training
and control sets, which satisfy $|A_0| \leq 2 m_0$. This means that
the ANNs are forced to at least partially extrapolate, rather than
interpolate, when estimating these errors. Recall also that values of
$|A_0|$ significantly larger than $2 m_0$ often lead to problems with
the calculation of the spectrum.

In spite of these caveats, we consider the overall performance of our
ANNs to be quite satisfactory. Already with $\unit[10]{fb^{-1}}$ of
simulated data the gaugino mass parameter $m_{1/2}$ can be determined
with a relative accuracy of $1$ to $2.5\%$. If $m_0$ is large enough
to significantly affect the squark masses (benchmark points 2 and 3),
it can be determined with a relative accuracy of about $4.5\%$; if $m_0$
is very small, as in point 1, this relative accuracy deteriorates.

Meaningful determinations of $A_0$ and $\tan\beta$ will need more
data. For the simulated measurements with $\unit[500]{fb^{-1}}$, the
estimated uncertainty on $\tan\beta$ varies between $4\%$ for point 4,
which had a large input value of $\tan\beta$, and $25\%$ for point 3,
which, as we had seen before, has the largest normalized control error
for this quantity. The estimated error on $A_0$ is roughly $25$ to
$35\%$ of the input value of $m_0$.

Finally, Table~\ref{tab:errors} also lists the correlation
coefficients. We see that most correlations are quite weak. The true
correlation coefficients should be independent of the luminosity, but
our estimates of these coefficients should, and do, fluctuate when the
(simulated) data set is increased.

We nevertheless observe consistently negative correlations between the
extracted values of $m_0$ and $m_{1/2}$ for benchmarks points 2, 3 and
4. This can be explained from the observation that increasing either
$m_{1/2}$ or $m_0$ will increase the masses of strongly interacting
superparticles, which will lead to a reduction of the total event
rate, and to an increase of the average $H_T$ values. To some extent
an increase in $m_0$ can therefore be compensated by a reduction of
$m_{1/2}$, and vice versa. This correlation is essentially absent for
benchmark point 1, which has $m_0^2 \ll m_{1/2}^2$ so that even the
squark masses are essentially independent of $m_0$. Similarly, the
mild positive correlation between $A_0$ and $m_{1/2}$ can be explained
from the observation that the coefficients $d_i$ in eq.(\ref{equ:mi})
are negative, while the $b_i$ are positive. The RG effect on the
scalar masses of increasing $A_0$ can therefore be compensated by
increasing $m_{1/2}$.


\begin{figure*}[!h]
\centering
\resizebox{1.\textwidth}{!}{
  \includegraphics{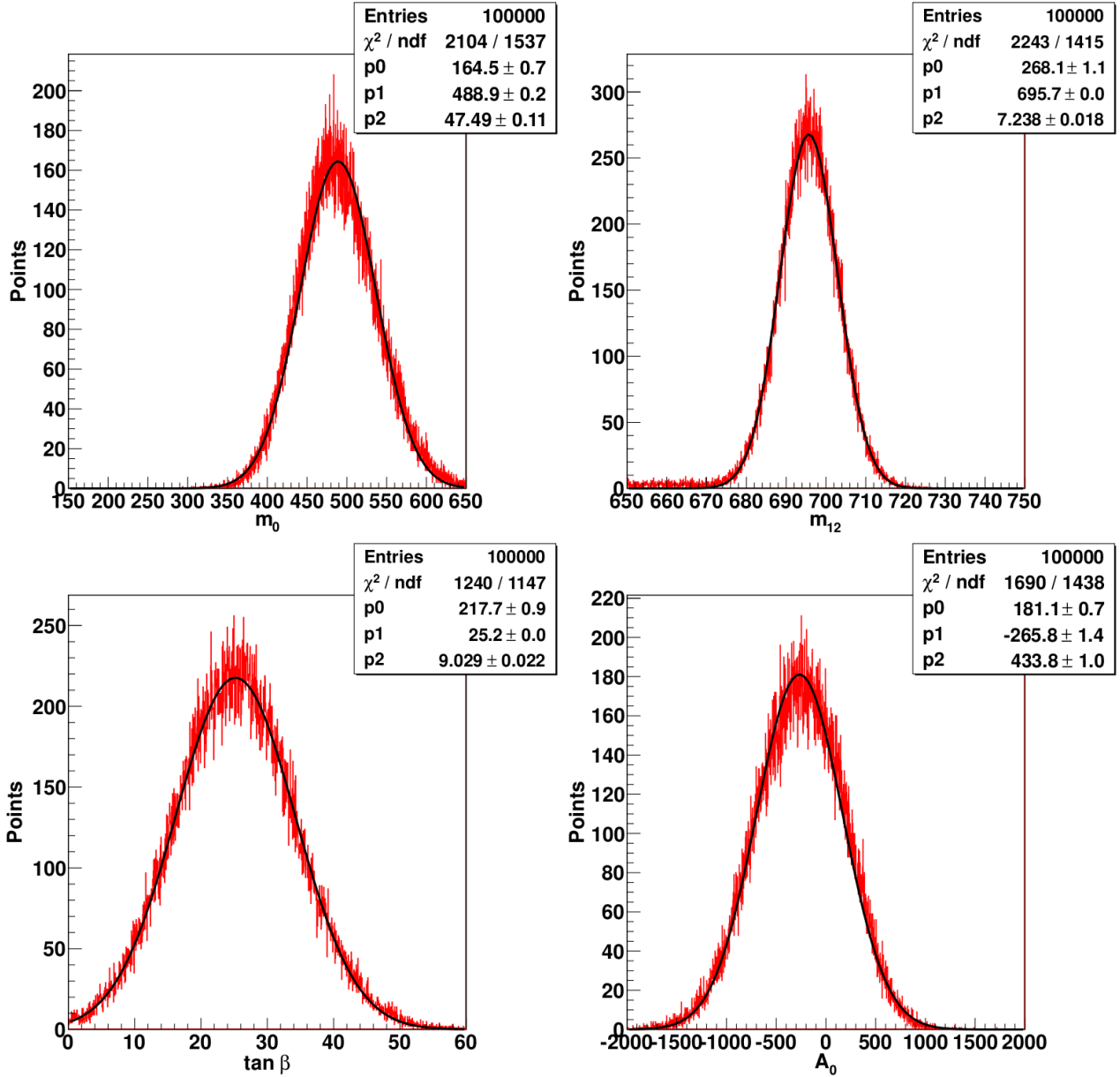}
}
\caption{The histograms show one--dimensional distributions of ANN
  outputs for benchmark point 4, obtained by feeding the neural
  networks with $100,000$ Gaussian distributed versions of the
  $\unit[10]{fb^{-1}}$ measurement. The dimensionful parameters $m_0$,
  $m_{1/2}$ and $A_0$ are given in GeV. The distributions are fitted
  to Gaussians (solid curves) of the form $g(x) = p_0 \cdot \exp(-1/2
  [(x - p_1)/p_2]^2)$ with $x$ being the appropriate CMSSM parameter,
  $p_1$ the mean value and $p_2$ the standard deviation. The input
  values are in the center of the $x$ axis of the respective plot. The
  neural network settings are given in Table~\ref{tab:performance}.}
\label{fig:1d_10}
\end{figure*}

As already noted, the standard deviations and correlation coefficients
listed in Table~\ref{tab:errors} have been computed using error
propagation. We conclude this Section by comparing these with the
results obtained by feeding Gaussian distributed variants of the
original ``measurements'' into our ANNs, as described in the first
part of Sec.~\ref{sec:Error}. We do this for benchmark point 4;
results for the other benchmark points are similar.\footnote{Some
  results for benchmark point 3 have already been shown in
  Sec.~\ref{sec:Error}.} 

Fig.~\ref{fig:1d_10} shows the distribution of the output of the four
ANNs for the simulated measurement with $\unit[10]{fb^{-1}}$ of
data. We note first of all that the binned distributions of ANN
outputs do indeed look rather Gaussian already for this smaller data
sample -- much more so than the corresponding distributions obtained
via $\chi^2$ minimization, see Fig.~\ref{fig:Fehler4}. Moreover, in
case of $m_0, \, m_{1/2}$ and $\tan\beta$ both the mean values of the
Gaussians (the $p_1$ values given in the inserts in the figure), which
are the final parameters estimates of the ANN derived in this manner,
and their widths (the $p_2$ values), which are the final estimates for
the uncertainty of these parameters, agree quite well with the results
listed in Table~\ref{tab:errors}. In case of the mean values, the two
methods yield estimates that agree to about $0.1$ estimated standard
deviations. The two estimates for the standard deviations differ
slightly more. This is not unexpected, since the standard deviation
estimated via error propagation itself has an error, which is not
negligible for this small (simulated) data set. Recall also that we
set all observables to zero that have been obtained from fewer than
$10$ events. Benchmark point 4 has $7$ events in class 9 and $13$
events in class 4. Using the method of error propagation, the
observables $O_{9,i}$ are thus ignored, while $O_{4,i}$ are
included. However, quite a few of the Gaussian distributed variants of
this measurement will have $\geq 10$ events in class 9 and/or $\leq 9$
events in class 4. These variants will thus feed {\em qualitatively}
different input in the ANNs than the original simulated measurement.

\begin{figure*}[!h]
\centering
\resizebox{1.\textwidth}{!}{
  \includegraphics{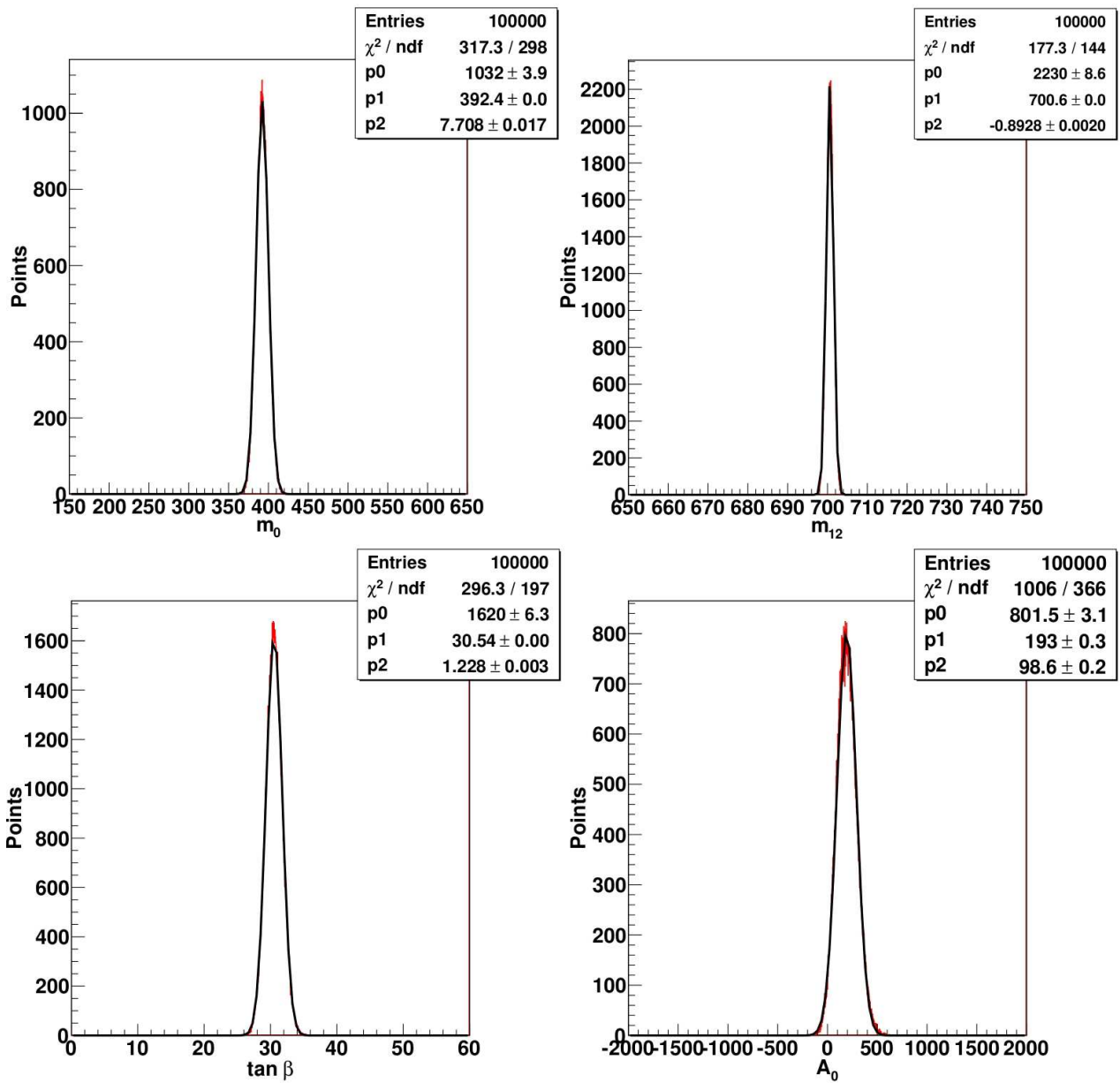}
}
\caption{As in Fig.~\ref{fig:1d_10}, but using $\unit[500]{fb^{-1}}$
  instead of a $\unit[10]{fb^{-1}}$ of simulated data.}
\label{fig:1d_500}
\end{figure*}

Both the statistical error of the estimated standard deviation, and
the systematic difference between the ANN inputs from different
variants of the same simulated measurement, are expected to decrease
with increasing luminosity. For example, for an integrated luminosity
of $\unit[500]{fb^{-1}}$, benchmark point 4 has $350$ events in class
9, which is about $8$ standard deviations away from the lower bound of
$500$ events now required for the inclusion of observables $O_{9,i}$;
we can therefore be quite certain that none of the up to $1,000,000$
variants of the simulated measurements has sufficiently many events of
class 9 for the $O_{9,i}$ to be included. We therefore expect the
differences between the two methods for determining the final ANN
output, and its estimated uncertainty, to agree better for higher
luminosity. This is confirmed by Fig.~\ref{fig:1d_500}. In particular,
the two estimates of the uncertainties now agree to better than $3\%$
in all cases.

\begin{figure*}[h!]
\centering
\resizebox{0.76\textwidth}{!}{
\includegraphics{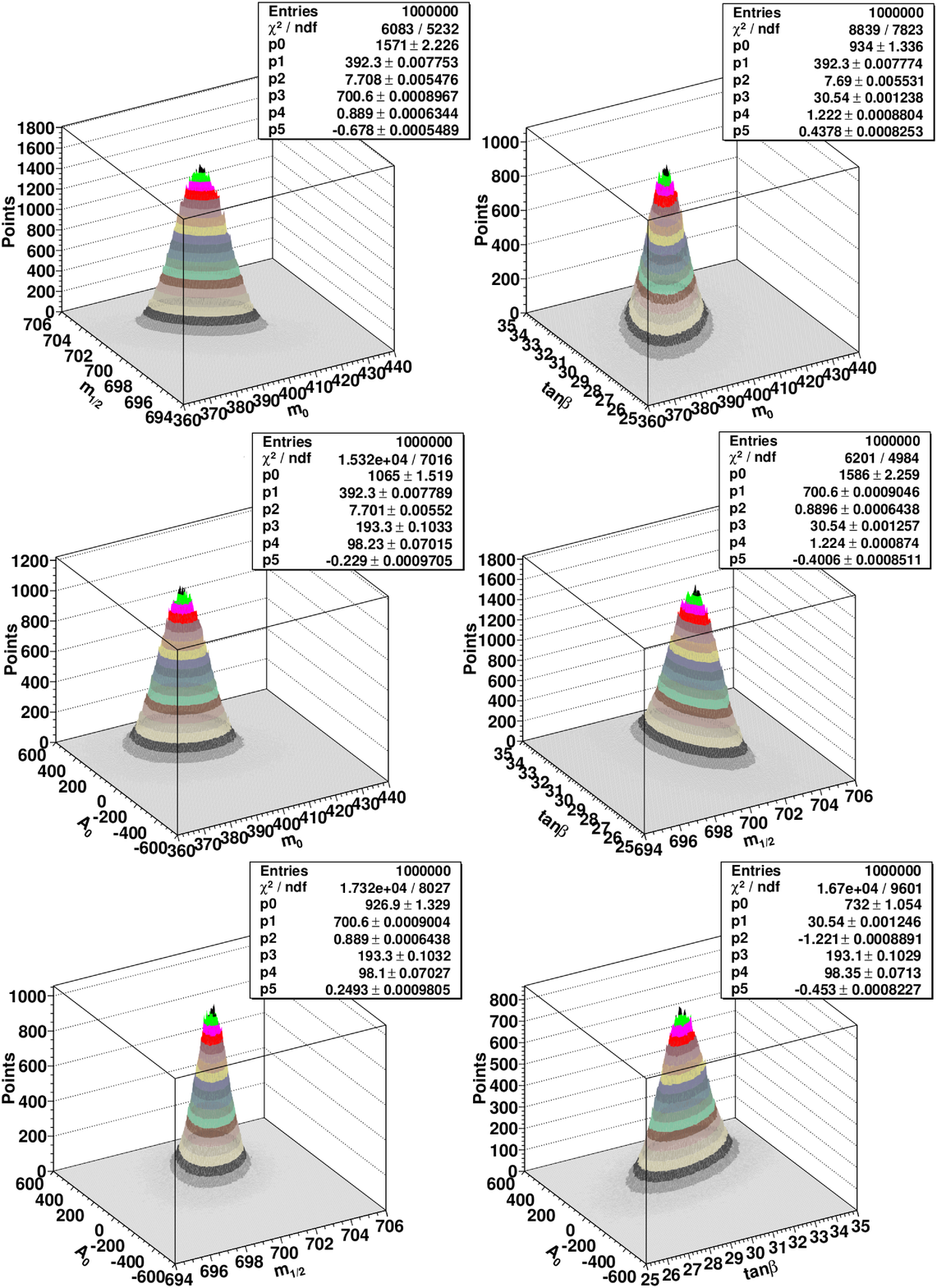}
}
\caption{Two--dimensional distributions of all pairs of CMSSM
  parameters as predicted from our ANNs for benchmark point 4, when
  fed with $1,000,000$ Gaussian distributed versions of the original
  simulated measurement, assuming an integrated luminosity of
  $\unit[500]{fb^{-1}}$. The dimensionful quantities $m_0$, $m_{1/2}$
  and $A_0$ are given in GeV. These distributions are fitted to
  Gaussians of the form $g(x, \, y) = p_0 \cdot \exp[-0.5/(1 - p_5^2)
  \cdot ([(x - p_1)/p_2]^2 + [(y - p_3)/p_4]^2 - 2 \cdot p_5/(p_2 \,
  p_4) \cdot (x - p_1) \cdot (y - p_3))]$, with $x$ and $y$ being the
  appropriate CMSSM parameters, $p_1$ and $p_2$ the mean value and
  standard deviation of $x$, $p_3$ and $p_4$ the mean value and
  standard deviation of $y$, and $p_5$ the correlation coefficient of
  eq.(\ref{equ:CorrelationCoefficient}).}
\label{fig:2d_500}
\end{figure*}

Recall that we used $100$ Gaussian distributed variants of ANN inputs
for each CMSSM training parameter set. Without this trick, the
distributions of the ANN outputs for $\tan\beta$ and $A_0$ would look
much less Gaussian, in particular for the smaller luminosity of
$\unit[10]{fb^{-1}}$. Moreover, the estimated uncertainties on the extracted CMSSM parameters would be
$\sim 20\%$ larger. The ANNs therefore clearly profit from the
information on the relative errors of our observables. Recall that
these additional training sets were obtained without additional event
generation. On the other hand, the time needed for training the ANNs
scales linearly with the size of the training sets.

Finally, Fig.~\ref{fig:2d_500} shows that Gaussian fits to
two--dimensional distributions of ANN outputs yield almost the same
estimates of the standard deviations as the one--dimensional fits or
the method of error propagation. These two--dimensional fits also
allow to determine the correlation coefficients (the $p_5$ values
given in the inserts in the figures). Again we observe quite close
agreement with the results derived from Gaussian error propagation.

Above we found some evidence that the errors from the
$\unit[500]{fb^{-1}}$ ``measurements'' are underestimated, since they
do not agree the ``theory'' error from the finite MC statistics used
in the training and control sets. Since Figs.~\ref{fig:1d_500} and
\ref{fig:2d_500} have been obtained with fixed ANNs, only varying the
input (i.e., the results of the ``measurements''), they do not reflect
this ``theory error'', either. It is nevertheless reassuring that two
methods which are computationally independent yield very similar
results.


\section{Summary and Conclusions}
\label{sec:Summary}

In this paper we investigated methods to determine the values of
underlying parameters from (simulated) measurements at the LHC, with
heavy emphasis on counting observables. Mostly for reasons of
computational simplicity, we did this for the CMSSM where only four
free parameters need to be determined; the sign of $\mu$ was fixed to
be positive.

Evidently the choice of observables is crucial. Here we used the same
observables as in ref.\cite{bod4}, where we had shown that they
perform well when trying to distinguish different (generalized) MSSM
parameter sets using a $\chi^2$ criterion. We also used the same cuts
as in ref.\cite{bod4}, even though the sparticle mass scale of the
benchmark points we used for the present analysis, which lie just
outside the currently excluded region, is significantly higher than in
our previous study. Moreover, in this proof--of--concept analysis we
ignored Standard Model backgrounds as well as statistical
uncertainties.

Our main result is that artificial neural networks (ANNs) can be used
for the determination of model parameters, including statistically
reliable estimates of their errors. In contrast, an in principle
straightforward $\chi^2$ minimization did not yield reliable results,
probably because we do not have the computational resources required
for the calculation of theoretical predictions for sufficiently many
different sets of model parameters with sufficiently small Monte Carlo
uncertainty. Moreover, the estimated errors on $m_0$ and $m_{1/2}$
from the $\chi^2$ minimization were about two to three times larger
than those obtained from the ANNs. We thus conclude that ANNs can yield
better and more reliable results with less computational effort than a
$\chi^2$ minimization.

Of course, the training of the ANNs was also affected by the finite MC
statistics used for deriving our theoretical predictions. However,
here the main requirement is that the error due to finite MC
statistics should be (much) smaller than the statistical error of the
(actual or simulated) measurement. In contrast, any ``smart'' algorithm
attempting to minimize a $\chi^2$ will need reliable information on
systematic changes of $\chi^2$ when the CMSSM parameters are changed
by relatively small amounts. The uncertainty on $\chi^2$ due to MC
statistics therefore needs to be (much) smaller than the (typically
quite small) change of $\chi^2$ induced by this small variation of the
parameters. 

Mathematically, an ANN is a function mapping (typically a rather large
number of) input values into one or more output value(s). In the case
at hand, the inputs are the 84 observables described in
Sec.~\ref{sec:Observables}. These observables have quite different
statistical uncertainties. This should affect the weights given to
these inputs. We took this into account in two ways. First, we simply
set all observables to zero that have been obtained from fewer than
$1$ event per fb$^{-1}$ of (simulated) data. Since we assume an
integrated luminosity of at least $\unit[10]{fb^{-1}}$, this
effectively removes very ``noisy'' observables. 

Second, for each training set of CMSSM parameters, we
generated $100$ variants where the observables were drawn from
multi--dimensional Gaussians, whose central values and covariances
were taken from the original simulation. For each set of CMSSM
parameters that we simulated for the training sets, the ANNs were
therefore confronted with $100$ slightly different sets of inputs
(observables) yielding the same outputs (CMSSM parameters). The ANNs
could thus learn the relative accuracy between the various
observables, which, at least for Gaussian statistics, should be
independent of the integrated luminosity. The trained ANNs could thus
be used on (simulated) data sets of any luminosity, as long as the
statistical uncertainty of this ``measurement'' is (much) larger than
the Monte Carlo uncertainty of the predictions in the training and
control sets. 

Since the training of the ANNs is independent of actual data, it can
be done before the measurement. Once actual measurements exist, the
ANN results for the CMSSM parameters could then be obtained with
negligible computational effort. In contrast, a $\chi^2$ minimization
has to be (re--)done for each measurement.

As expected the CMSSM parameters $m_0$ (scalar mass parameter) and
$m_{1/2}$ (gaugino mass parameter) could be determined relativeley
well for all four benchmark points already with an integrated
luminosity of $\unit[10]{fb^{-1}}$. In the best cases these could be
determined to $\unit[4.5]{\%}$ for $m_0$ and $\unit[1]{\%}$ for
$m_{1/2}$. With this luminosity, leading to around $1,000$ events
after cuts, $\tan\beta$ and $A_0$ could at best be determined very roughly.

On the other hand, with a luminosity of $\unit[500]{fb^{-1}}$,
$m_{1/2}$ can be determined with statistical uncertainty well below
$1\%$. The statistical error on $m_0$ then amounts to $5$ to
$\unit[10]{GeV}$ for our four benchmark points. For three of the
points, $\tan\beta$ could be determined with an error of $\pm 2$ or
better. Finally, the error on $A_0$ was about $25$ to $35\%$ of the
input value of $m_0$. We also computed the full covariance matrix, and
found that most correlations are quite weak.

These results were obtained with two different methods. In one method,
the central values were obtained by simply feeding the simulated
measurements into the trained ANNs, and the uncertainties and
correlation coefficients were computed using Gaussian error
propagation. Alternatively we generated numerous Gaussian distributed
variants of the original simulated measurements, and fitted Gaussian
distributions to the outputs of the ANNs. The results of these two
methods agreed quite well.

One disadvantage of using ANNs is that they do not automatically give a measure for the goodness of the fit (at most only an indication by the shapes of the Gaussian output distributions): even if Nature is described by a completely different theory, the ANNs will output some values of the free parameters of the (wrong) theory on which they have been trained when confronted with actual data. One will have to simulate the assumed theory with the values of the free parameters determined by the ANNs in order to determine the quality of the fit, e.g.~by computing the $\chi^2$. However, the numerical effort required for this is trivial compared to the effort required for the determination of the values of the free parameters.

Our results can be improved in a number of ways. First of all, we
ignored all information on the Higgs sector. In the context of the
MSSM, knowledge of the mass of one of the CP--even neutral Higgs
bosons will greatly reduce the allowed parameter space. One can also
try to tag top quarks \cite{toptag} or Higgs bosons \cite{higgstag} in
the final state using subjet techniques, or to devise dedicated sets
of cuts that attempt to isolate specific decay chains. Moreover,
kinematic features (edges or kinks) could be included. All this would
increase the number of inputs fed into the ANNs, and could thus
increase their ability to determine the underlying parameters.
Conversely it might be possible to remove some of the observables from
the list of input parameters without significant loss of information.
Since our algorithm should automatically assign low weights to
observables with little discriminating power, this would presumably
not improve the performance of the ANNs very much, but it could reduce
the computational effort. Similarly, the number of Gaussian
distributed variants generated for each training set of CMSSM
parameters could perhaps be reduced without degrading the performance;
the time needed to train the ANNs is essentially proportional to this
number. Finally, we did not consider ANNs with two (or more) layers of
hidden neurons; this more complicated architecture might allow to
reduce the total number of hidden neurons, and perhaps also the total
number of weights that need to be determined, which would speed up the
training process.

However, before trying to further optimize the performance of the ANNs
one should make the set--up more realistic, by including Standard
Model backgrounds as well as systematic uncertainties. The cuts could
then be optimized for each benchmark point separately, as e.g.~done in
\cite{reach,baer}. This would not increase the computational effort, since
we already use different ANNs for the different benchmark points
(or regions). Systematic uncertainties could be introduced as in
\cite{bod4}. Moreover, if the method is applied to supersymmetric
scenarios, one should consider benchmark points that have a Higgs
boson of the correct mass and coupling, in agreement with recent data
\cite{higgs_dis}. In the framework of the CMSSM this is known to push
the squark mass scale to quite large values \cite{cmssm_higgs}; this
will presumably greatly reduce the possibility to extract
$m_0$. However, if scalar masses are not required to unify, first and
second generation scalars could still have masses similar to those in
our benchmark scenarios. Alternatively, one could introduce additional
Higgs superfields to increase the mass of the lightest CP--even Higgs
boson, as e.g.~in the NMSSM \cite{nmssm}.  

The purpose of this paper was to show that artificial neural networks
do have the potential to determine quantitatively the values of the
parameters of the underlying theory, and the corresponding
(statistical) uncertainties. Further, more realistic studies are thus
well worth the effort.

\subsection*{Acknowledgments}

NB wishes to thank the ``Bonn-Cologne Graduate School of Physics and
Astronomy'' for financial support. This work was partially
supported by the BMBF--Theorieverbund and by the Helmholtz Alliance
``Physics at the Terascale''.

\begin{appendix}
\end{appendix}
\section{Calculation of Weights}
\setcounter{footnote}{0}
\label{sec:CalculationOfWeights}

In this Appendix we discuss the calculation of the weights of a neural
network with $84 + 1$ input neurons, $v + 1$ hidden neurons and one
output neuron; this describes the ANNs we constructed in
Sec.~\ref{sec:NeuralNetwork}. Since there is only one output neuron,
we suppress the index $r=1$ on the corresponding weights in
eq.(\ref{equ:outputNeuronInput}).

We begin by combining all weights in one vector $\vec{w}$. The first
$85 \cdot v$ entries are weights in the first weight layer, and the
remaining $v + 1$ entries are from the second weight layer:
\begin{eqnarray} \label{ea0}
\vec{w} & = & \left( w_1, \, \dots, \, w_W \right)^T  \nonumber \\
	& = & \left( w_1, \, \dots, \, w_{v \cdot 85}, \,
          w_{v \cdot 85 + 1}, \, \dots, \, w_W \right)^T \nonumber
        \\ 
 	& = & \left( w^{(1)}_{10}, \, w^{(1)}_{11}, \, \dots, \,
          w^{(1)}_{1 \, 84}, \, w^{(1)}_{20}, \, \dots, \, w^{(1)}_{v
            \, 84}, \, w^{(2)}_0, \, \dots, \, w^{(2)}_v
        \right)^T \, .
\end{eqnarray}
In total the weight vector thus has $W = 85 \cdot v + v + 1 = 86 \cdot
v + 1$ entries. The $w_m, \ m = 1, \dots, 85 \cdot v$ are related
to the $w^{(1)}_{ai}, \ a = 1, \dots, v, \ i = 0, \dots, 84$ via $w_m =
w^{(1)}_{ai}$ with $a = \lceil m/85 \rceil$\footnote{$\lceil x \rceil$
  means the smallest integer that is larger than or equal to the real
  number $x$; e.g.~$\lceil 1/6 \rceil = 1 = \lceil 1 \rceil$.} and $i
= [(m - 1) \mod 85]$. Similarly, for $n = 85 \cdot v + 1, \dots W$ we have
$w_n = w^{(2)}_b$ with $b = (n - 1 - v \cdot 85)$.

During the training of the ANN the weight vector is adjusted
iteratively.  We denote the weight vector in learning step $t$ by
$\vec{w}_t$.

\subsection*{First Step}

As mentioned in Sec.~\ref{sec:Initialization} the first weight vector
$\vec{w}_1$ is chosen randomly from a Gaussian distribution. In the
first improvement, $\vec{w}$ is changed in direction $\vec{s}_1 =
-\vec{g}_1$ equal to the negative gradient $\vec{g}$. The first
gradient vector $\vec{g}_1 = \vec{g}(\vec{w}_1) = \vec{\nabla}
F(\vec{w}_1)$ is calculated using the function $F$ describing the error
the ANN makes in reproducing the training set. We wish to minimize
this function of the weights. Since the location of the minimum is
independent of the normalization, we use the simple definition
\begin{equation} \label{equ:errorFunction}
F = \sum\limits_{\ell = 1}^{N} F^\ell = \sum\limits_{\ell = 1}^{N} \frac{1}{2}
\left( y^\ell(\vec{x}^\ell) - k^\ell \right)^2. 
\end{equation}
As in Sec.~\ref{sec:NeuralNetwork} $N$ is the number of training sets;
in our applications, $N \simeq 100,000$. Again as in
Sec.~\ref{sec:NeuralNetwork}, $y^\ell(\vec{x}^\ell)$ is the output
that the ANN computes from the normalized input $\vec{x}^\ell$ of
training set $\ell$, while the correct (inversely) normalized CMSSM
output of the training set is labeled $k^\ell$; the normalization of
the input and output has been described in
Sec.~\ref{sec:Normalization}. The first $85 \cdot v$ entries of the
gradient vector can be computed from eqs.(\ref{equ:errorFunction}),
(\ref{equ:hiddenNeuronInput}) and (\ref{equ:outputNeuronInput}):
\begin{equation} \label{ea1}
\vec{\nabla}_m F(\vec{w}) = \frac{\partial F}{\partial w_m} =
\frac{\partial F}{\partial w^{(1)}_{ai}} = \sum\limits_{\ell = 1}^{N}
\delta^{(1) \ell}_a \, x^\ell_i \,.
\end{equation}
The remaining $v + 1$ entries are:
\begin{equation} \label{ea2}
\vec{\nabla}_n F(\vec{w}) = \frac{\partial F}{\partial w_n} =
\frac{\partial F}{\partial w^{(2)}_b} = \sum\limits_{\ell = 1}^{N}
\delta^{(2) \ell} \, h(z^\ell_b) \, . 
\end{equation}
Here, $h$ is the hidden neuron processing function, $h(z^\ell_b) =
\tanh(z^\ell_b)$, and $z_b^\ell = \sum_{i=0}^{84} w^{(1)}_{bi} x^\ell_i$
as in eq.(3.1). Moreover, $\delta^{(1) \ell}_a$ in eq.(\ref{ea1}),
with $a = 1, \, \dots, \, v$, and $\delta^{(2) \ell}$ in
eq.(\ref{ea2}) are abbreviations for:
\begin{equation} \label{equ:delta1}
\delta_a^{(1) \ell} =  \frac{\partial F^\ell}{\partial
  y^\ell} \, \frac{\partial y^\ell}{\partial z^\ell_a} =
  \delta^{(2) \ell} \, w_a^{(2)} \cdot h^\prime(z^\ell_a) \,,
\end{equation}
where $h'$ stands for the derivative of $h$, i.e.~$h^\prime(z^\ell_b)
= (1 - \tanh^2(z^\ell_b))$, and
\begin{equation} \label{equ:delta2}
\delta^{(2) \ell} = \frac{\partial F^\ell}{\partial y^\ell} = y^\ell - k^\ell\,.
\end{equation}

\subsection*{Repeated Steps}

The following steps of the calculation steps are repeated until one is
satisfied that the (global) minimum of the normalized control error
(\ref{equ:NormalizedError}) has been reached. The new weight vector
is calculated in step $t \geq 2$ from the expression
\begin{equation} \label{equ:newWeights}
\vec{w}_{t} = \vec{w}_{t-1} + \alpha_t \, \vec{s}_{t-1} \,.
\end{equation}
The calculation of $\vec{s}_1 = - \vec{g}_1 = - \vec{\nabla}
F(\vec{w}_1)$ has already been described in eqs.(\ref{ea1}) to
(\ref{equ:delta2}). The coefficient $\alpha_t$ can be computed
from the Hessian matrix $H_{t-1}$:
\begin{equation}
\alpha_t = - \frac{\vec{s}_{t-1}^T \, \vec{g}_{t-1} }
  {\vec{s}_{t-1}^T \, H_{t-1} \,   \vec{s}_{t-1} } \, . 
\end{equation}
The explicit calculation of the Hessian matrix is shown at the end of
this Appendix. With the new weights the normalized control error of
eq.(\ref{equ:NormalizedError}) can be calculated and depending
on the stopping criterion the learning process might be terminated.

If the stopping criterion is not fulfilled the new gradient vector
$\vec{g}_t$ would be calculated, just as in eqs.(\ref{ea1}) to
(\ref{equ:delta2}). Next, the new search direction is computed from
\begin{equation}
\vec{s}_t = -\vec{g}_t + \beta_t \, \vec{s}_{t-1} \, ,
\end{equation}
where the coefficient $\beta_t$ is also calculated from the Hessian
matrix:
\begin{equation}
\beta_t = \frac{\vec{g}_t^T \, H_{t-1} \, \vec{s}_{t-1}} {\vec{s}_{t-1}^T \,
  H_{t-1} \, \vec{s}_{t-1}} \, . 
\end{equation}
Now the next step $t \rightarrow t + 1$ can be taken, starting with
the calculation of the new weights from eq.(\ref{equ:newWeights}). 

\subsection*{Hessian Matrix}

The main numerical effort in the training process is the repeated
calculation of the Hessian matrix. This is a symmetric $W \times W$
matrix, i.e.~its dimension is determined by the number of hidden
neurons.  It is given by the matrix of second derivatives of the error
function $F$ with respect to the weights:
\begin{equation}
H_{mn} = \frac{\partial^2 F}{\partial w_m \partial w_n}\, .
\end{equation}
Its numerical value will in general be different for each step during
the training process. For our architecture, with one layer of hidden
neurons, we can distinguish three different cases: both weights are
from the first layer; both weights are from the second layer; or one
weight is from the first and the other from the second weight layer:
\begin{itemize}
\item[1)] Both weights are from the first weight layer, i.e.~$m, \, n
  = 1, \, \dots, \, 85 \cdot v$: 
\begin{eqnarray*}
H_{mn} & = & \sum\limits_{\ell = 1}^{N} \frac{\partial^2 F^\ell}{\partial
  w^{(1)}_{ai} \partial w^{(1)}_{bj}} = \sum\limits_{\ell = 1}^{N}
\frac{\partial}{\partial w^{(1)}_{ai}} (\delta^{(1) \ell}_b \, x^\ell_j) \\ 
      & = & \sum\limits_{\ell = 1}^{N} \frac{\partial}{\partial
             w^{(1)}_{ai}} \left[ \delta^{(2)\ell} \, w_b^{(2)} \cdot
             h^\prime(z^\ell_b) \right] \, x^\ell_j \\ 
      & = & \sum\limits_{\ell = 1}^{N} \left[ w_a^{(2)} \, w_b^{(2)} \,
             h^\prime(z^\ell_a) \, h^\prime(z^\ell_b) + \delta_{ba} \,
             \delta^{(2)\ell} \, w_b^{(2)} \, h^{\prime\prime}(z^\ell_b) \right]
           \, x^\ell_i \, x^\ell_j \,.
\end{eqnarray*}
Here we have used eqs.(\ref{equ:errorFunction}), (\ref{equ:delta1}),
(\ref{equ:delta2}), (\ref{equ:hiddenNeuronInput}) and
(\ref{equ:outputNeuronInput}). The index pairs $ai$ and $bj$ are
computed from the indices $m$ and $n$ as described following
eq.(\ref{ea0}) above. The second derivative of the hidden neuron
processing function is
\begin{equation}
h^{\prime\prime}(z) = 2 \, \tanh(z) \, (\tanh^2(z) - 1) = 2 \, h(z) \,
(h^2(z) - 1) \, . 
\end{equation}
\item[2)] Both weights are from the second weight layer, i.e.~$ m, \,
  n = v \cdot 85 + 1, \, \dots, \, W $: 
\begin{eqnarray*}
H_{mn} & = & \sum\limits_{\ell = 1}^{N} \frac{\partial^2 F^\ell}{\partial
  w^{(2)}_a \partial w^{(2)}_b } = \sum\limits_{\ell = 1}^{N}
\frac{\partial}{\partial w^{(2)}_a } (\delta^{(2) \ell} \, h(z^\ell_b))
\\ 
      & = & \sum\limits_{\ell = 1}^{N} \, h(z^\ell_a) \, h(z^\ell_b) \,,
\end{eqnarray*}
where $a = m - 85 \cdot v - 1, \ b = n - 85 \cdot v - 1$.
\item[3)] One weight each from the first and the second weight layer,
  i.e.~$m = v \cdot 85 + 1, \, \dots, \, W$ and $n = 1, \, \dots,
  \, v \cdot 85 $ or vice versa: 
\begin{eqnarray*}
H_{mn} & = & \sum\limits_{\ell = 1}^{N} \frac{\partial^2 F^\ell}{\partial
  w^{(2)}_a \partial w^{(1)}_{bi}} \\ 
      & = & \sum\limits_{\ell = 1}^{N} \frac{\partial}{\partial
        w^{(2)}_a} \left[ \delta^{(2) \ell} \, w_b^{(2)} \cdot
        h^\prime(z^\ell_b) \right] \, x^\ell_i \\ 
      & = & \sum\limits_{\ell = 1}^{N} \left( h(z^\ell_a) \, w_b^{(2)} +
        \delta_{ba} \, \delta^{(2)\ell} \right) \, h^\prime(z^\ell_b) \,
      x^\ell_i 
\end{eqnarray*}
\end{itemize}

\end{document}